%% file: main.tex
\documentclass[cernpreprint,texlive=2016,txfonts,UKenglish]{atlasdoc} 

\usepackage{atlaspackage}
\usepackage{atlasphysics}

\usepackage{main-defs}
\usepackage{subfig} 
\usepackage{dcolumn}
\usepackage{epstopdf,epsfig}

\hypersetup{pdftitle={ATLAS draft},pdfauthor={The ATLAS Collaboration}}

\input{commands}
\input{main-metadata}

\begin{document}

\tableofcontents

\input{intro}

\input{detector}

\input{objects}

\input{data_presel}

\input{bkg_modeling}

\input{measurement}

\input{results}

\FloatBarrier

\input{summary}

\section*{Acknowledgements}

\input{Acknowledgements}

\FloatBarrier

\sloppy 
\printbibliography

\newpage 
\input{atlas_authlist}

\end{document}

%% file: commands.tex
\newcommand{\ljets}{\ensuremath{\ell{}+\text{jets}}}
\newcommand{\ejets}{$e$+jets}
\newcommand{\mujets}{$\mu{}$+jets}

\newcommand{\lumi}{\ensuremath{20.3~\ifb{}}}

\newcommand{\herwig}{{\textsc Herwig}}

\newcommand{\pythia}{{\textsc Pythia}}

\newcommand{\mcnlo}{{\textsc Mc@nlo}}

\newcommand{\wjets}{$W$+jets}

\newcommand{\zjets}{$Z$+jets}

\newcommand{\AC}{\ensuremath{A_{\text{C}}}}

\newcommand{\mtt}{\ensuremath{m_{\ttbar{}}}}
\newcommand{\pttt}{\ensuremath{p_{\text{T},\ttbar{}}}}
\newcommand{\ytt}{\ensuremath{y_{\ttbar{}}}}

\newcommand{\betatt}{\ensuremath{\beta_{z,\ttbar{}}}}
\newcommand{\pp}{\ensuremath{pp}}

\providecommand{\abs}[1]{\lvert#1\rvert} 
\newcommand{\dy}{\ensuremath{\Delta{}\abs{y}}}

\newcommand{\vect}[1]{\boldsymbol{#1}}
\newcommand{\thetavec}{\ensuremath{\vect{\theta{}}}}
\newcommand{\Truth}{\ensuremath{\vect{T}}}
\newcommand{\Data}{\ensuremath{\vect{D}}}
\newcommand{\Bckg}{\ensuremath{\vect{B}}}
\newcommand{\Reco}{\ensuremath{\vect{R}}}
\newcommand{\TrasfMatrix}{\ensuremath{\mathcal{M}}}

\newcommand{\conditionalProb}[2]{\ensuremath{p\left(#1|#2\right)}}
\newcommand{\conditionalLhood}[2]{\ensuremath{\mathcal{L}\left(#1|#2\right)}}
\newcommand{\prior}{\ensuremath{\pi{}\left(\Truth{}\right)}}

\let\oldmarginpar\marginpar
\renewcommand\marginpar[1]{
  \-\oldmarginpar[\raggedleft\footnotesize #1]{\raggedright\footnotesize #1}
}

%% file: main-metadata.tex
\AtlasTitle{ Measurement of the charge asymmetry in top-quark pair 
  production in the lepton-plus-jets final state 
  in \ensuremath{pp} collision data at $\sqrt{s}=8\TeV{}$ with the
  ATLAS detector }

\author{The ATLAS Collaboration}

\AtlasRefCode{TOP-2014-16}

\PreprintIdNumber{CERN-PH-EP-2015-217}

\AtlasJournal{EPJC}

\AtlasAbstract{
 This paper reports inclusive and differential measurements of the \ttbar{}
  charge asymmetry \AC{} in \lumi{} of $\sqrt{s} = 8~\TeV{}$ \pp{}
  collisions recorded by the ATLAS experiment at the Large Hadron Collider at CERN. Three
  differential measurements are performed as a function of 
  the invariant mass, transverse momentum and longitudinal boost of the \ttbar{} system.
  The \ttbar{} pairs are selected in the single-lepton channels
  ($e$ or $\mu$) with at least four jets, and a
  likelihood fit is used to reconstruct the \ttbar{} event
  kinematics. A Bayesian unfolding procedure  is performed 
  to infer the asymmetry at parton level from the observed data distribution. 
  The inclusive \ttbar{}  charge asymmetry is measured to be
  $\AC{}  =  0.009 \pm 0.005$ (stat.$+$syst.). 
  The inclusive and differential measurements are compatible with the values predicted
  by the Standard Model.

}

%% file: intro.tex
\section{Introduction}
\label{sec:intro}

The 8~\TeV{} proton--proton ($pp$) collision data delivered by the CERN Large Hadron Collider (LHC) represents a unique
laboratory for precision measurements of the top-quark properties.
One interesting feature of \ttbar{} production is the difference in rapidity between top quarks and top antiquarks. 
In $pp$ collisions, this distinct behaviour of top quarks and antiquarks is called 
the charge asymmetry, \AC{} (defined in Eq.~(\ref{Ac}) below). The Standard Model (SM) expectation computed at 
next-to-leading order (NLO) in quantum chromodynamics (QCD), including electroweak corrections, predicts \AC{} to be at the one percent level~\cite{Bern:2012}.
Previous asymmetry measurements at the LHC by both the CMS and ATLAS 
collaborations based on the 7~\TeV{} data, and by the CMS 
collaboration based on the 8~\TeV{} data, do not report any significant deviation 
from the SM predictions~\cite{CMS_ljets,CMS_ljets2,Aad:2013cea,Aad:2015jfa,Khachatryan:2015mna,Khachatryan:2015oga}.
Charge asymmetry measurements are largely limited by the size of the available data sample, 
and therefore the larger dataset recorded by the ATLAS detector at $\sqrt{s}$ = 8~\TeV{} allows for 
an improvement on the precision of the measurement from the $\sqrt{s}$= 7~TeV{} dataset.

At hadron colliders, \ttbar{} production is predicted to be symmetric under the 
exchange of top quark and antiquark at leading order (LO).
At NLO, the process 
$q\bar{q}\to t\bar{t}g$ develops an asymmetry in the top-quark rapidity distributions, 
due to interference between processes with initial- and final-state gluon emission. 
The interference between the Born and the NLO diagrams of the $q\bar{q}\to t\bar{t}$ process 
also produces an asymmetry.
The $qg\to\ttbar{}g$ production process is also asymmetric, but its contribution is much smaller than that from \qqbar{}.

In $q\bar{q}$ scattering processes in 
$p\bar{p}$ collisions at the Tevatron, the direction of the incoming quark almost always coincides with 
that of the proton, and this knowledge of the direction of the incoming quarks allows one to define a direct measurement 
of the forward-backward asymmetry, $A_{\mathrm{FB}}$~\cite{Aguilar-Saavedra:2014kpa, Czakon:2014xsa, CDF1,D01}. 
In $pp$ collisions at the LHC,
since the colliding beams are symmetric, it is not possible to use the direction of the incoming quark 
to define an asymmetry.
However, valence quarks carry on average a larger fraction of the proton momentum than sea antiquarks,
hence top quarks are more forward and top antiquarks are more central.
Using this feature it is possible to define a forward--central asymmetry for  
the \ttbar{} production, referred to as the charge asymmetry, \AC{}~\cite{Aguilar-Saavedra:2014kpa,Jung:2011zv,Diener:2009ee} :
\begin{linenomath}
\begin{equation}
\AC{} = \frac{N(\dy{}>0) - N(\dy{} <0)}{N(\dy{} >0) + N(\dy{} <0)},
\label{Ac}
\end{equation}
\end{linenomath}
where $\dy{} \equiv |y_t| - |y_{\bar{t}}| $  is the difference between the absolute value of the top-quark rapidity $|y_t|$ and the absolute 
value of the top-antiquark rapidity $|y_{\bar{t}}|$.
At the LHC, the dominant mechanism for \ttbar{} production is the gluon 
fusion process, while production via the \qqbar{} or the $qg$ interactions is small. 
Since ${gg \to t\bar{t}}$ processes are charge-symmetric, they only contribute 
to the denominator of Eq.~(\ref{Ac}), thus diluting the asymmetry.

Several processes beyond the Standard Model (BSM) can alter \AC{}~\cite{Jung:2011zv, AXI,Djouadi:2009nb, KK, Jung:2009jz, Shu:2009xf, JA:2011, 
AguilarSaavedra:2011hz, Dorsner:2009mq, Grinstein:2011yv, Ligeti:2011vt, Ferrario:2009bz, Frampton:2009rk}, 
either with anomalous vector or axial-vector couplings (e.g. axigluons) or via interference with SM processes. 
Different models also predict different asymmetries as a function of the invariant mass \mtt{}, the transverse momentum 
\pttt{} and the longitudinal boost \betatt{} along the $z$-axis\footnote{ATLAS 
uses a right-handed coordinate system with its origin at the nominal interaction point (IP) in the 
centre of the detector and the $z$-axis coinciding with the axis of the beam pipe. The $x$-axis points from
the IP to the centre of the LHC ring, and the $y$-axis points upward. Cylindrical coordinates ($r$,$\phi$) are used 
in the transverse plane, $\phi$ being the azimuthal angle around the beam pipe.}  of the \ttbar{} system~\cite{AguilarSaavedra:2011ci}. 
The interest in precisely measuring charge asymmetries in top-quark pair production at the LHC has grown after the CDF and D0 
collaborations reported measurements of $A_{\mathrm{FB}}$ that were 
significantly larger than the SM predictions, in both the inclusive and differential 
case as a function of \mtt{} and of the rapidity of the \ttbar{} system, \ytt{}~\cite{CDF1,CDF2,CDF3,Abazov:2015fna,Abazov:2014cca,D01}.
For the most general BSM scenarios~\cite{AguilarSaavedra:2012prl}, the \AC{} measurements from the LHC are 
still compatible with the Tevatron results.  However, for specific simple models~\cite{AguilarSaavedra:2011hz}, 
tension still exists between the LHC and Tevatron results.
This motivates the interest in a more precise measurement of the \ttbar{} production charge asymmetry at the LHC.

In this paper, a measurement of the \ttbar{} production charge asymmetry in the single-lepton final state is reported. 
To allow for comparisons with theory calculations, a Bayesian unfolding procedure is applied to account for distortions due to the acceptance 
and detector effects, leading to parton-level \AC{} measurements.
The data sample at a centre-of-mass energy of 8~\TeV{}, corresponding to an integrated luminosity 
of $20.3~\ifb$~\cite{Aad:2013ucp}, is used to measure \AC{} inclusively and differentially as a function of \mtt{}, \pttt{} and \betatt{}. 

This paper is organised as follows. The ATLAS detector is introduced in Sect.~\ref{sec:detector}, followed by the object
reconstruction in Sect.~\ref{sec:object_reco} and the event selection in Sect.~\ref{sec:data_presel}. 
The signal and background modelling is described in Sect.~\ref{sec:bkg_model} and the procedure to measure $\AC{}$ in
Sect.~\ref{sec:acmeas}.
Finally, the results are presented and interpreted in Sect.~\ref{sec:result}, followed by the conclusions in 
Sect.~\ref{sec:conclusion}.

%% file: detector.tex
\section{ATLAS detector}
\label{sec:detector}

The ATLAS detector~\cite{atlas-detector} consists of the following
 main subsystems: an inner tracking system 
immersed in a 2 T magnetic field provided by a superconducting solenoid,
electromagnetic (EM) and hadronic calorimeters, and a muon spectrometer 
incorporating three large superconducting toroid magnets composed of eight coils each.
The inner detector (ID) is composed of three subsystems: the pixel detector, the semiconductor tracker  
and the transition radiation tracker. The ID provides tracking information
in the pseudorapidity\footnote{The pseudorapidity is defined in terms of the polar angle $\theta$ as 
$\eta = - \ln \tan(\theta/2)$ and transverse momentum and energy are defined relative to the beam line as 
$\pt{}= p \sin \theta$ and $E_{\mathrm T} =E \sin \theta$. The angular distances are given in terms of
$\Delta R= \sqrt{(\Delta \eta)^2 + (\Delta \phi{})^2} $, where $\phi{}$ is the azimuthal angle around the beam pipe.} 
range $|\eta|<2.5$, calorimeters measure energy deposits (clusters) for $|\eta|<4.9$, and the muon spectrometer records tracks within $|\eta|<2.7$.
A three-level trigger system~\cite{atlas-trigger-2010} is used to select interesting events.
It consists of a level-1 hardware trigger, reducing the event rate to at most 75 kHz,
followed by two software-based trigger levels, collectively referred to as the high-level trigger, yielding a recorded event rate of 
approximately 400~Hz on average, depending on the data-taking conditions.

%% file: objects.tex
\section{Object reconstruction}
\label{sec:object_reco}

This measurement makes use of reconstructed electrons, muons, jets, $b$-jets and missing transverse momentum. 
A brief summary of the main reconstruction and identification criteria applied 
for each of these objects is given below. 

Electron candidates are reconstructed from clusters in the 
EM calorimeter that are matched to reconstructed tracks in the inner
detector.  They are required to have a transverse energy, $\et$, greater than $25~\GeV{}$ 
and $|\eta_{\mathrm cluster}| < 2.47$, where $\eta_{\mathrm cluster}$ is the pseudorapidity 
of the electromagnetic energy cluster in the calorimeter with respect
to the geometric centre of the detector.  
Candidates are required to satisfy the {\textsl tight} quality requirements~\cite{Aad:2014fxa} 
and are excluded if reconstructed in the transition region between the barrel and endcap 
sections of the EM calorimeter, $1.37 < |\eta_{\mathrm cluster}| < 1.52$.
They are also required to originate less than 2 mm along the $z$-axis 
(longitudinal impact parameter) from the
selected event primary vertex (PV)\footnote{The method of selecting the PV is described in Sect.~\ref{sec:data_presel}.} and to satisfy two isolation criteria. 
The first one is calorimeter-based and consists of a requirement
on the transverse energy sum of cells within a cone of size 
$\Delta R = 0.2$ around the electron direction. The second one is 
a track-based isolation requirement made on the track transverse momentum ($\pt$) sum around the
electron in a cone of size $\Delta R = 0.3$. In both
cases, the contribution from the electron itself is excluded and the isolation cuts
are optimised to individually result in a 90\%  
efficiency for prompt electrons from
$Z\rightarrow e^+e^-$ decays.

Muon candidates~\cite{Aad:2014zya,Aad:2014rra} are reconstructed using the combined 
information from the muon spectrometer and the inner detector.  They are required to satisfy
$\pt > 25~\GeV{}$ and $|\eta|<2.5$ and analogously to electrons, the muon track longitudinal 
impact parameter with respect to the PV is required to be less than 2 mm.
Muons are required to satisfy a $\pt$-dependent track-based isolation:
the scalar sum of the track $\pt$ within a cone of 
variable size around the muon, $\Delta R =10~\GeV{}/\pt^\mu$ 
(excluding the muon track itself) must be less than 5\% of the muon \pt{} ($\pt^\mu$), corresponding to a 
97\% selection efficiency for prompt muons from $Z\rightarrow \mu^+\mu^-$ decays.

Jets are reconstructed with the anti-$k_t$ 
algorithm~\cite{Cacciari:2008gp,Cacciari:2005hq,Cacciari:2011ma} with a
radius parameter $R=0.4$ from calibrated topological
clusters~\cite{atlas-detector} built from energy deposits in the
calorimeters.  Prior to jet finding, a local cluster calibration scheme~\cite{Cojocaru:2004jk,LCW2}
is applied to correct the topological cluster energies for the effects of 
the  noncompensating response of the calorimeter, 
dead material and out-of-cluster leakage. The corrections are obtained
from simulations of charged and neutral particles and validated with data. 
After energy calibration~\cite{Aad:2014bia}, jets are required to have 
$\pt > 25~\GeV{}$ and $|\eta| < 2.5$.  
Jets from additional simultaneous $pp$ interactions (pileup) are suppressed by requiring that the absolute value of the 
jet vertex fraction (JVF)\footnote{The jet vertex fraction is defined as the fraction 
of the total transverse momentum of the jet's associated tracks that is contributed 
by tracks from the PV.} 
for candidates with $\pt<50~\GeV{}$ and $|\eta|<2.4$ is above 0.5~\cite{ATLAS-CONF-2013-083}.
All high-$\pt$ electrons are also reconstructed as jets, so 
the closest jet within $\Delta R=$ 0.2 of a selected electron 
is discarded to avoid double counting of electrons as jets.
Finally, if selected electrons or muons lie within $\Delta R$ = 0.4 of selected jets, 
they are discarded.

Jets are identified as originating from the hadronisation of a $b$-quark 
($b$-tagged) via an algorithm that uses 
multivariate techniques to combine information from the impact
parameters of displaced tracks as well as topological properties of
secondary and tertiary decay vertices reconstructed within the jet~\cite{BTaggingEfficiency,CLTaggingEfficiency}.
The algorithm's operating point used for this measurement
corresponds to 70\% efficiency to tag $b$-quark jets, a 
rejection factor for light-quark and gluon jets of $\sim$130 and
a rejection factor of $\sim$5 for $c$-quark jets, as determined for 
jets with $\pt >20~\GeV$ and $|\eta|<2.5$ in simulated $\ttbar$ events.

The missing transverse momentum 
(with magnitude $\met$) is constructed from the negative vector sum of all calorimeter energy deposits~\cite{Aad:2012re}.
The ones contained in topological clusters are calibrated at 
the energy scale of the associated high-$\pt$ object (e.g.~jet or electron).
The topological cluster energies are corrected using
the local cluster calibration scheme discussed in the jet reconstruction paragraph above. 
The remaining contributions to the $\met$ are called unclustered energy.
In addition, the \met{} calculation 
includes contributions from 
the selected muons, and muon energy deposits in the calorimeter are removed 
to avoid double counting.

%% file: data_presel.tex
\section{Event selection}
\label{sec:data_presel}

Only events recorded with an isolated or non-isolated single-electron or single-muon trigger 
under stable beam conditions with all detector subsystems operational are considered.

The triggers have thresholds on $p^\ell_{\mathrm T}$, the transverse momentum (energy) of the muon (electron).  
These thresholds are 24~\GeV{}  for isolated single-lepton triggers and 60 (36)~\GeV{} for non-isolated single-electron (single-muon) triggers. 
Events satisfying the trigger selection are required to have at least one 
reconstructed vertex with at least five associated tracks of $\pt > 400$ MeV, consistent with originating 
from the beam collision region in the $x$--$y$
plane. If more than one vertex is found, the hard-scatter PV is taken
to be the one which has the largest sum of the squared transverse momenta
of its associated tracks. 

Events are required to have exactly one candidate electron or muon
and at least four jets satisfying the quality and kinematic criteria discussed in Sect.~\ref{sec:object_reco}. 
The selected lepton is required to match, with $\Delta R < 0.15$, the lepton reconstructed by the 
high-level trigger.
Events with additional electrons satisfying a looser identification criteria 
based on a likelihood variable~\cite{ATLAS:2014wga} 
are rejected in order to suppress 
di-leptonic backgrounds (\ttbar{} or \zjets{}). 
At this point, the events are separated into three signal regions defined by the
number of $b$-tagged jets  (zero, one and at least two).
  
In order to further suppress multijet and \zjets{} backgrounds in events with exactly zero or one $b$-tagged jets, 
the following requirements on \met{} and \mtw{}\footnote{$\mtw = \sqrt{2 p^\ell_{\mathrm T} \met (1-\cos\Delta\phi)}$, where
$p^\ell_{\mathrm T}$  is the transverse momentum (energy) of the muon (electron) and $\Delta\phi$ is the
azimuthal angle separation between the lepton and the direction of
the missing transverse momentum.} are applied: $\mtw{} + \met{} > 60~\GeV{}$ for events with exactly zero or one $b$-tagged jets, and
$\met{} > 40~(20)~\GeV{}$ for events with exactly zero (one) $b$-tagged jets.

After the event selection, the main background is the production of $W$+jets events.
Small contributions arise from multijet, single top quark, $Z$+jets and diboson
($WW,WZ,ZZ$) production. 
For events with exactly one (at least two) $b$-tagged jet(s), 216465 (193418) data events are observed, of which 68\% (89\%) 
are expected to be \ttbar{}.

%% file: bkg_modeling.tex
\section{Signal and background modelling}
\label{sec:bkg_model}

Monte Carlo simulated samples are used to model the \ttbar{} signal and all 
backgrounds except for those from multijet events, which are estimated from data.
All simulated samples utilise {\textsc Photos} (version 2.15)~\cite{Golonka:2005pn} to simulate 
photon radiation and {\textsc Tauola} (version 1.20)~\cite{Jadach:1990mz} to simulate $\tau$ decays.
They also include simultaneous $pp$ interactions (pile-up), generated using {\textsc Pythia} 8.1~\cite{Sjostrand:2007gs}, and reweighted 
to the number of interactions per bunch crossing in data (on average 21 in 2012). 
Most of them are processed through a full {\textsc Geant4}~\cite{AGO-0301-2} simulation of the detector response~\cite{Aad:2010ah}, and 
only the alternative \ttbar{} samples described in Sect.~\ref{sec:ttbarmod} are produced using the ATLAS 
fast simulation that employs parameterised showers in the calorimeters~\cite{ATLASAFII}.
Finally, the simulated events are reconstructed using the same software as the data.
Further details on the modelling of the signal and each of the backgrounds are provided below.

\subsection{$\ttbar$ signal}
\label{sec:ttbarmod}
The default simulated \ttbar{} events are generated with the NLO generator 
{\textsc Powheg-Box} (version 1, r2330)~\cite{Nason:2004rx,Frixione:2007vw,Alioli:2010xd} using the CT10 PDF 
set~\cite{Lai:2010vv} interfaced to {\textsc Pythia} (version 6.427)~\cite{Sjostrand:2006za} with the CTEQ6L1 PDF set and the 
Perugia2011C set of tunable parameters (tune)~\cite{Skands:2010ak} for the underlying event (UE). The 
$h_{\mathrm{damp}}$ factor, which is the model parameter that controls matrix element/parton shower matching 
in {\textsc Powheg-Box} and effectively regulates the high-$\pt$ radiation, is set to the top-quark mass.

The alternative samples used to study the modelling of \ttbar{} are:
\begin{itemize}

\item  \mcnlo{} (version 4.01)~\cite{Frixione:2002ik}
using the CT10 PDF set and interfaced to \herwig{} (version 6.520)~\cite{Corcella:2000bw} and {\textsc Jimmy} (version 4.31)~\cite{Butterworth:1996zw}.

\item {\textsc Powheg-Box} using the CT10 PDF and setting the $h_{\mathrm{damp}}$ parameter to infinity,
interfaced to {\textsc Pythia} (version 6.426) with the CTEQ6L1 PDF set and the 
Perugia2011C UE tune.  

\item {\textsc Powheg-Box} using the CT10 PDF and setting the $h_{\mathrm{damp}}$ parameter to infinity, 
and interfaced to {\textsc Herwig} with the CTEQ6L1 PDF set and {\textsc Jimmy} to simulate the UE.

\item {\textsc AcerMC}~\cite{Kersevan:2004yg} using the CTEQ6L1 PDF set and interfaced to {\textsc Pythia} (version 6.426).
\end{itemize}

All \ttbar{}  samples are generated assuming a top-quark mass of $172.5~\GeV{}$ and are normalised 
to the theoretical cross section of $\sigma_{t\bar{t}}= 253^{+13}_{-15}$~pb  calculated at 
next-to-next-to-leading order (NNLO) in QCD including resummation of next-to-next-to-leading 
logarithmic (NNLL) soft gluon terms  with {\textsc Top++} 
v2.0~\cite{Cacciari:2011hy,Beneke:2011mq,Baernreuther:2012ws,Czakon:2012zr,Czakon:2012pz,Czakon:2013goa,Czakon:2011xx}.

\subsection{$W/Z$+jets background}
\label{sec:wzjets}

Samples of events with a $W$ or $Z$ boson produced in association with jets  ($W/Z$+jets) 
are generated with up to five additional partons using the {\textsc Alpgen} (version 2.14)~\cite{Mangano:2002ea} 
LO generator and the CTEQ6L1 PDF set, interfaced to {\textsc Pythia} (version 6.426) for parton showering and fragmentation.
To avoid double counting of partonic configurations generated by both the matrix-element calculation and the parton shower, 
a parton--jet matching scheme (``MLM matching'')~\cite{Mangano:2001xp} is employed. 
The $W$+jets samples are generated separately for $W$+light-jets, $Wb\bar{b}$+jets, $Wc\bar{c}$+jets, and $Wc$+jets. 
The $Z$+jets samples are generated separately for $Z$+light-jets, $Zb\bar{b}$+jets, and $Zc\bar{c}$+jets. Overlap between 
$W/ZQ\bar{Q}$+jets ($Q=b,c$)  events generated from the matrix-element calculation and those generated from parton-shower 
evolution in the $W/Z$+light-jets samples is avoided via an algorithm based on the angular separation between the extra heavy quarks:
if $\Delta R(Q,\bar{Q})>0.4$, the matrix-element prediction is used, otherwise the parton-shower prediction is used. 
The $Z$+jets background is normalised to its inclusive NNLO theoretical cross section~\cite{Melnikov:2006kv}, 
while data is used to normalise  $W$+jets (see below for details).
Further corrections are applied to $Z$+jets simulated events in order to better describe data in the preselected sample.
A correction to the heavy-flavour fraction was derived to reproduce the relative rates of $Z$+2-jets
events with zero and one $b$-tagged jets observed in data. In addition, the $Z$ boson $\pt$ spectrum was compared between
data and the simulation in $Z$+2-jets events, and a reweighting function was derived in order to improve the modelling as
described in Ref.~\cite{Aad:2015eua}.

The procedure to estimate the normalisation of the \wjets{} background in data
exploits the difference in production cross section at the LHC
between $W^+$ and $W^-$, where the $W^+$ production cross section is 
higher than $W^-$~\cite{Aad:2011dm}. This is due to the 
higher density of $u$ quarks in protons with respect to $d$ quarks,
which causes more $u\bar{d}\to W^+$ to be produced than $d\bar{u}\to W^-$. 
The $W$ boson charge asymmetry is then defined as the difference between the 
numbers of events with a single positive or negative lepton divided by the sum.
The prediction for the $W$ boson charge asymmetry in \wjets{} production is little 
affected by theoretical uncertainties and can be
exploited, in combination with constraints from $W^+$ and
$W^{-}$ data samples, to derive the correct overall normalisation 
for the MC sample prediction.
The $W$ boson charge asymmetry 
depends
on the flavour composition of the sample, as the size and sign of the
asymmetry varies for $Wb\bar{b}$+jets, $Wc\bar{c}$+jets, $Wc$+jets, and $W$+light-jets production. 
The in situ calibration procedure embedded in the unfolding and 
described in Sect.~\ref{sec:measurement}, uses 
different signal and control regions to determine the normalisation of the \wjets{} background.

\subsection{Multijet background}
Multijet events can enter the selected data sample through several
production and misreconstruction mechanisms.  In the electron
channel, the multijet background consists of non-prompt electrons from 
heavy-flavour decays or photon conversion
or jets with a high fraction of their energy deposited in the EM calorimeter.  
In the muon channel, the background contributed by multijet events is predominantly 
due to final states with non-prompt muons, such as those from semileptonic $b$-
or $c$-hadron decays.  
The multijet background normalisation and shape are estimated from data using the ``Matrix Method'' (MM)
technique.

The MM exploits differences in the properties used for lepton identification 
between prompt, isolated leptons from
$W$ and $Z$ boson decays (referred to as ``real leptons'') and those
where the leptons are either non-isolated or result from the
misidentification of photons or jets (referred to as ``fake leptons''). For this purpose, two samples
are defined after imposing the event selection described in Sect.~\ref{sec:data_presel},
differing only in the lepton identification criteria: a ``tight''
sample and a ``loose'' sample, the former being a subset of the
latter.  The tight selection employs the final lepton identification
criteria used in the analysis. For the loose selection, the lepton isolation
requirements are omitted for both the muon and electron channels, and the quality requirements 
are also loosened for the electron channel.
The method assumes that the number of
selected events in each sample ($\nl$ and $\nt$) can be expressed as a
linear combination of the numbers of events with real and fake
leptons, so that the number of multijet events in the tight sample 
is given by 
\begin{equation}
N^{\mathrm{tight}}_{\mathrm{multijet}} = \frac{\epsf}{\epsr-\epsf}(\epsr \nl - \nt)
\end{equation}
where $\epsr$ ($\epsf$) represents the probability for a real (fake) lepton that satisfies
the loose criteria to also satisfy the tight. Both of these probabilities are measured
in data control samples. 
To measure $\epsr$, samples enriched in real leptons from $W$ boson decays 
are selected by requiring high $\met$ or transverse mass $\mtw$. The average $\epsr$ is 0.75 (0.98) in the
electron (muon) channel. To measure $\epsf$, samples enriched in multijet background are selected
by requiring either low $\met$ (electron channel) or high transverse impact parameter significance for the lepton track (muon channel).
The average $\epsf$ value is 0.35 (0.20) in the electron (muon) channel. Dependencies 
of $\epsr$ and $\epsf$ on quantities such as lepton $\pt$ and $\eta$, $\Delta R$ between the
lepton and the closest jet, or number of $b$-tagged jets, are parameterised in order to obtain
a more accurate estimate.

\subsection{Other backgrounds}

Samples of single-top-quark backgrounds corresponding to the $t$-channel, $s$-channel and $Wt$ production mechanisms 
are generated with {\textsc Powheg-Box} (version 3.0)~\cite{Alioli:2009je,Re:2010bp} using the CT10 PDF set.  All samples are generated assuming a top-quark mass of 
$172.5~\GeV{}$ and are  interfaced  to {\textsc Pythia} (version 6.425) with the CTEQ6L1 PDF set and the Perugia2011C UE tune.  
Overlaps between the \ttbar\ and $Wt$ final states are removed using the ``diagram removal'' scheme~\cite{Frixione:2005vw}.
The single-top-quark samples are normalised to the approximate NNLO theoretical cross 
sections~\cite{Kidonakis:2011wy,Kidonakis:2010ux,Kidonakis:2010tc} using the MSTW 2008 NNLO PDF set. 

Most of the diboson $WW/WZ/ZZ$+jets samples are generated using {\textsc Alpgen} (version 2.13), 
with up to three additional partons, and using the 
CTEQ6L1 PDF set, interfaced to {\textsc Herwig} and {\textsc Jimmy} (version 4.31) for parton showering, fragmentation and UE modelling. 
For the $WW$+jets samples, it is required that at least one of the $W$ bosons decays leptonically, 
while for the $WZ/ZZ$+jets samples, it is demanded that at least one of the $Z$ bosons decays leptonically.
Additional samples of $WZ$+jets, requiring the $W$ and $Z$ bosons to decay
leptonically and hadronically, respectively, are generated with up to three additional partons, 
including massive $b$- and $c$-quarks, using {\textsc Sherpa} v1.4.1~\cite{Gleisberg:2008ta} and the CT10 PDF set.
All diboson samples are normalised to their NLO theoretical cross sections~\cite{Campbell:1999ah}.

%% file: measurement.tex
\section{Charge asymmetry measurement}
\label{sec:acmeas}

To measure the charge asymmetry in top-quark pair events, the full \ttbar{}
system is reconstructed (Sect.~\ref{sec:ttbareco}) and the \dy{} spectra 
are unfolded to measure parton-level charge asymmetries (Sect.~\ref{sec:unfolding}) using the 
estimation of the backgrounds and systematic uncertainties (Sect.~\ref{sec:systematics}). Significant 
improvements to the analysis method with respect to the 7~\TeV{} measurement~\cite{Aad:2013cea} 
have been made, and these improvements are detailed in the description of the measurement in Sect.~\ref{sec:measurement}.

\begin{figure}[!t]
\centering

\includegraphics[width=0.45\textwidth]{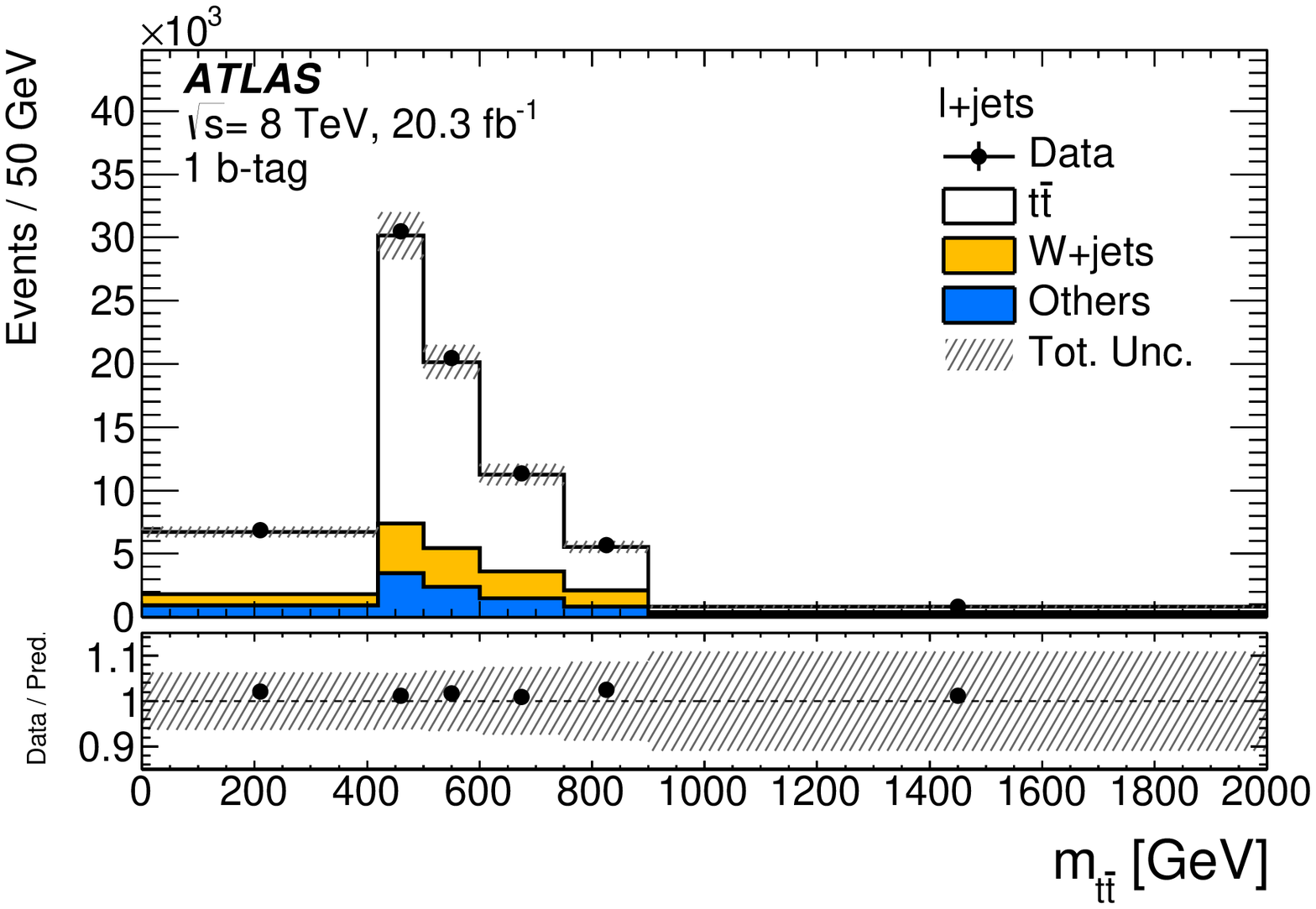}
\includegraphics[width=0.45\textwidth]{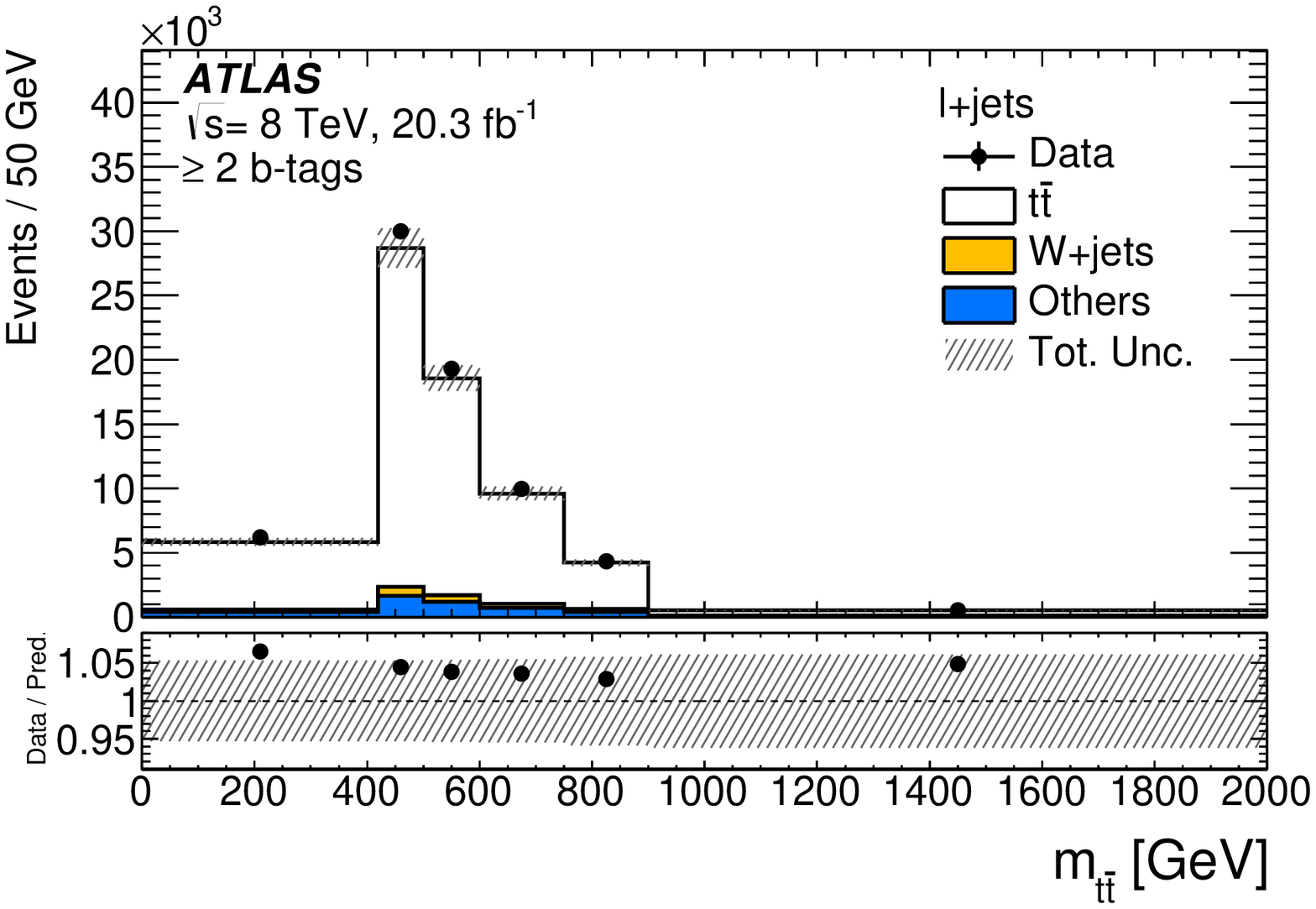}

\includegraphics[width=0.45\textwidth]{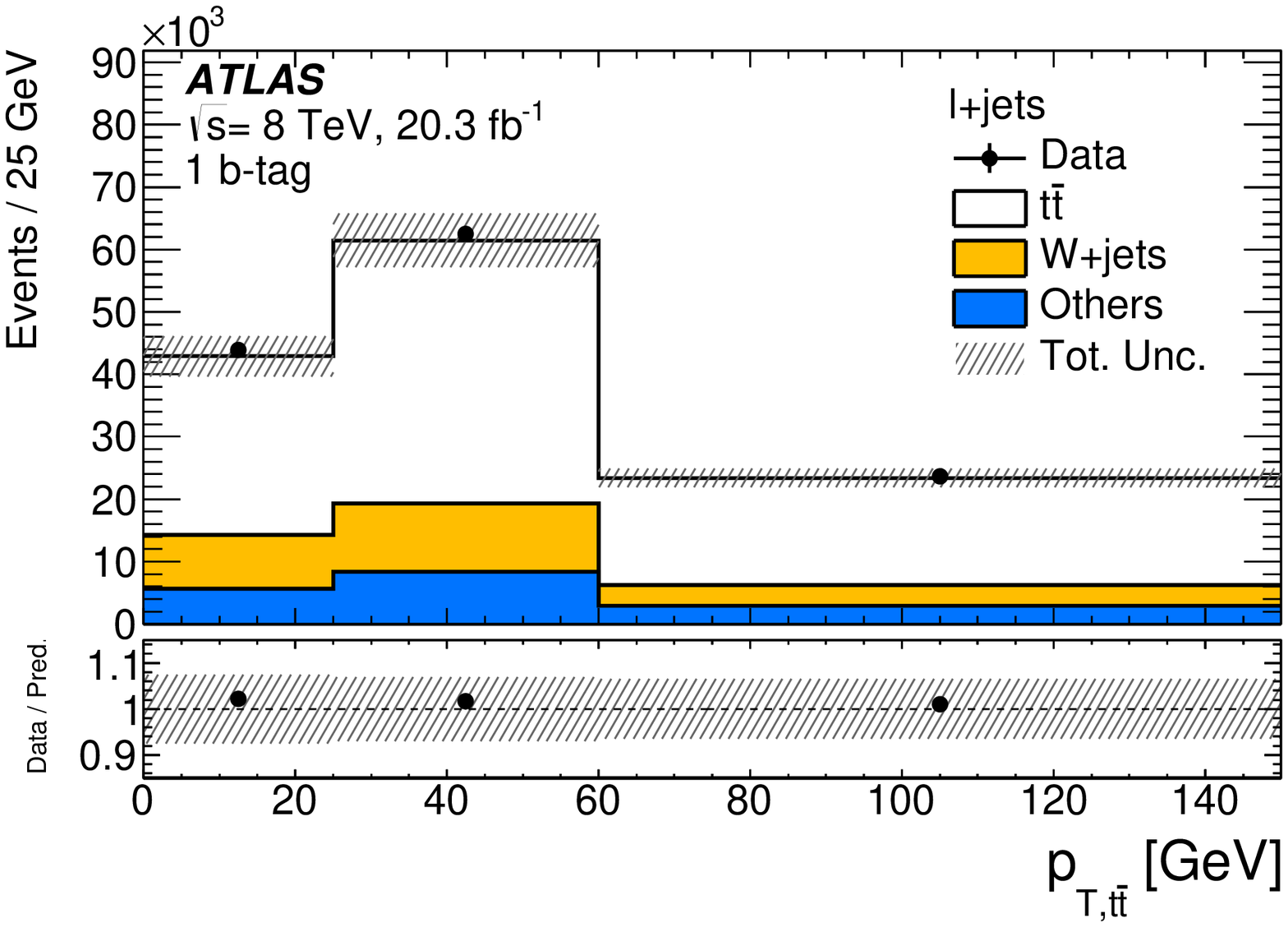}
\includegraphics[width=0.45\textwidth]{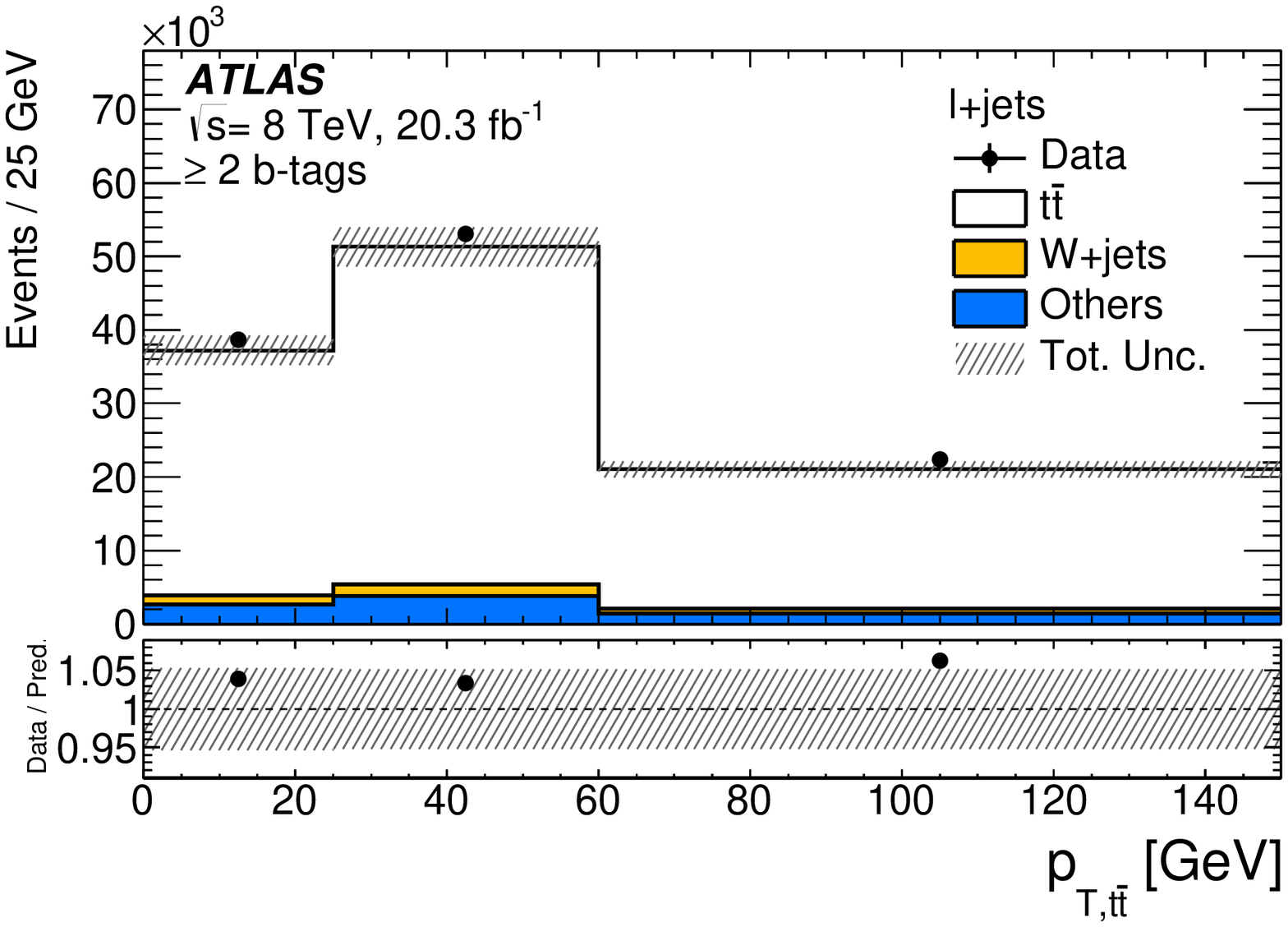}

\includegraphics[width=0.45\textwidth]{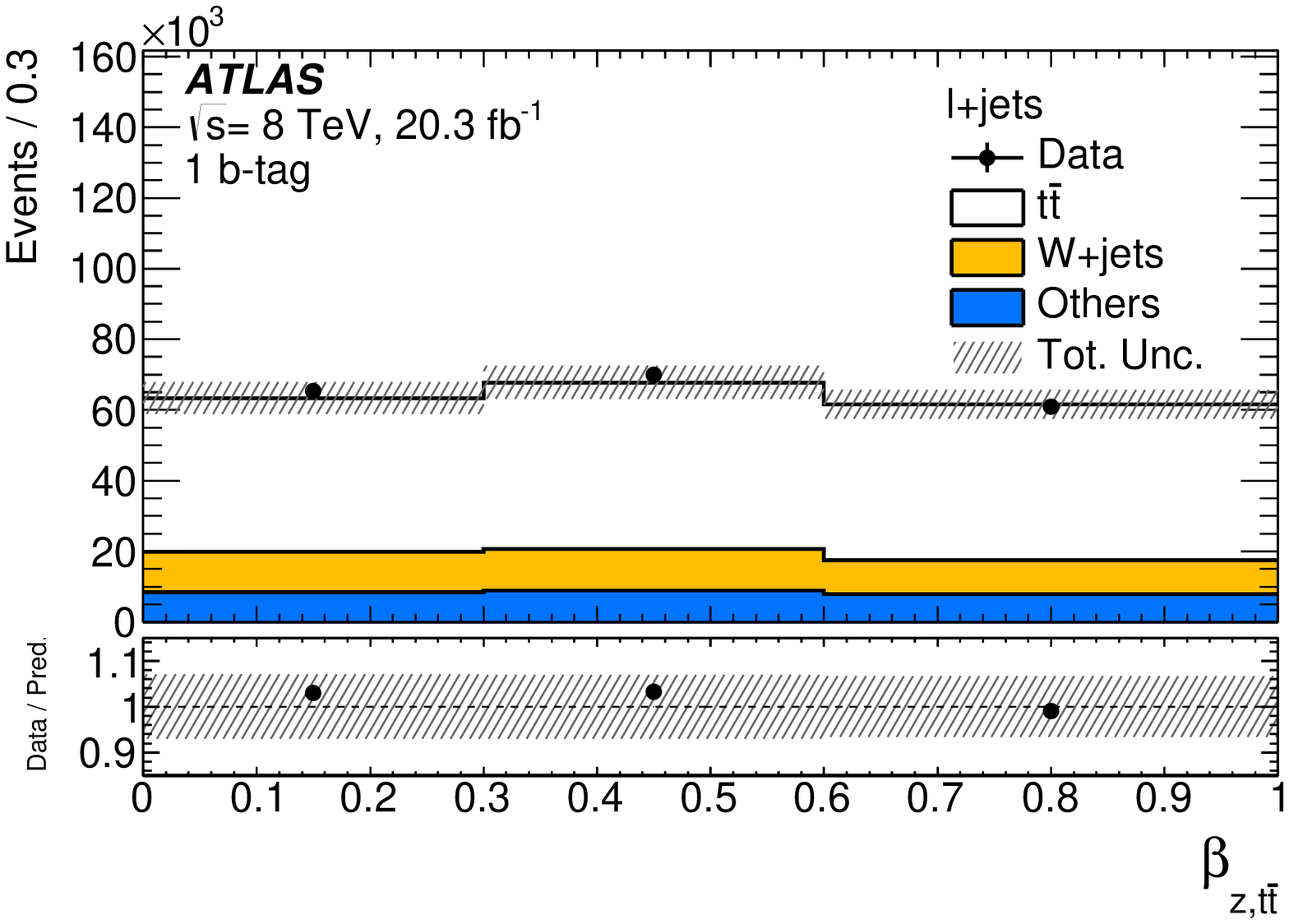}
\includegraphics[width=0.45\textwidth]{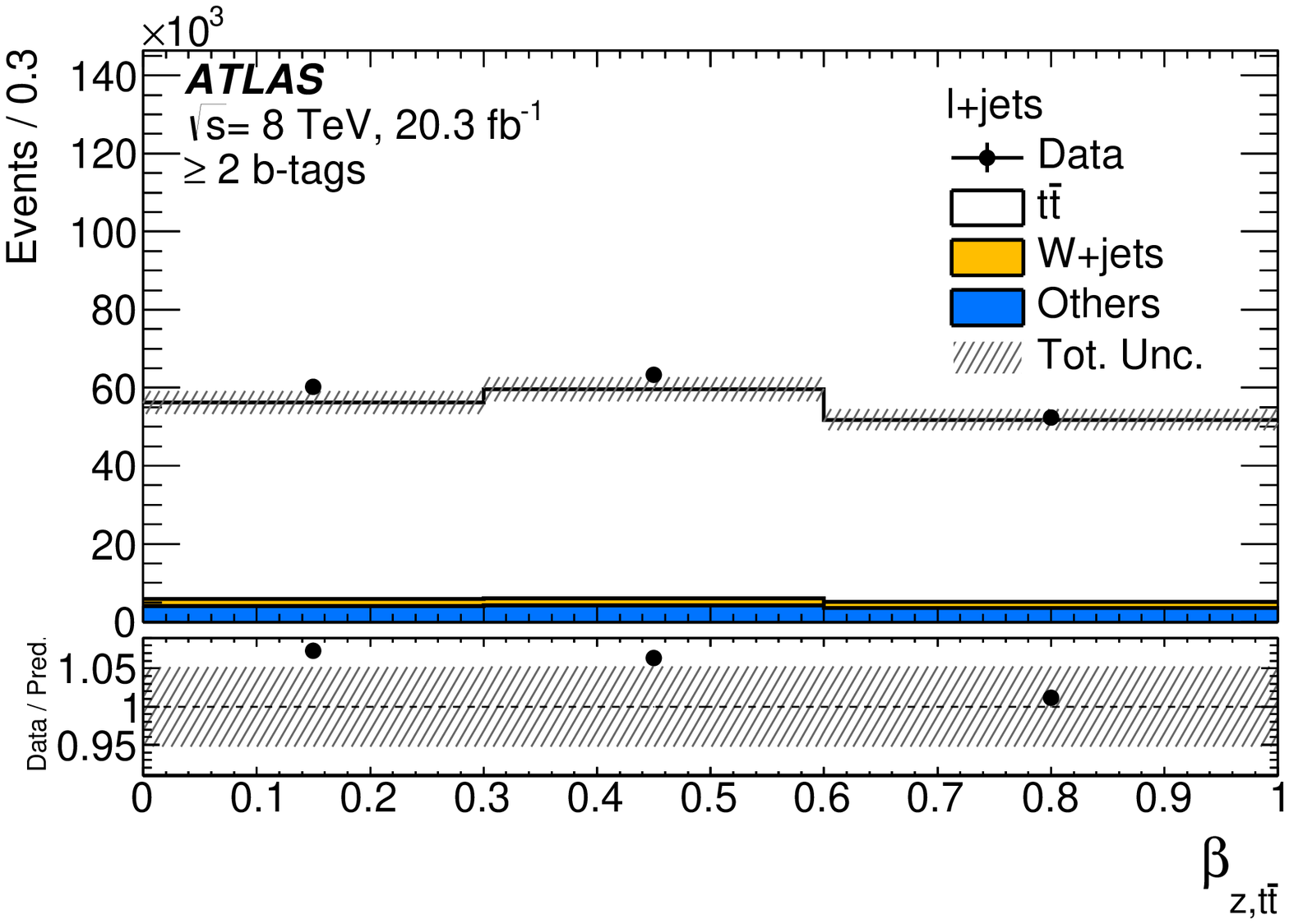}

\caption{Comparison between data and prediction for the \ejets{} and \mujets{} channels
combined for distributions of kinematic quantities,
in the sample with one $b$-tagged jet (left) and in the sample with at least two $b$-tagged jets (right).
From top to bottom: invariant mass \mtt{}, transverse momentum \pttt{}, $z$-component of
the velocity of the \ttbar{} system \betatt{}.
The total uncertainty, before the unfolding process, on the signal and background
estimation is shown together with statistical uncertainty as a black hashed band, and
the binnings are those that are used for the differential measurements.
The bottom part of each plot shows the ratio of the data to the predicted value together with
combined statistical and systematic uncertainties.}
\label{fig:datamcreco2012}
\end{figure}

\subsection{Reconstruction of the \ttbar{} kinematics}
\label{sec:ttbareco}
The reconstruction of the \ttbar{} system is achieved using 
a kinematic fit~\cite{KLFitterPaper}  that assesses the compatibility of the observed event
with the decays of a $\ttbar{}$ pair based on a likelihood
approach. 
The basic reconstruction method is explained in Ref.~\cite{ATLAS_ljets}, but some modifications 
are introduced as discussed in the following paragraph.

In events with four or five jets, all jets are considered in the fit.
For events where more than five jets are reconstructed, only
the two jets with the highest likelihood to be $b$-jets, according to the multivariate selection 
(see Sect.~\ref{sec:object_reco}),
and, of the remaining jets, the three with the highest $\pt$ are considered in the fit. 
This selection of input jets for the likelihood
was chosen to optimise the correct-sign fraction 
of reconstructed \dy{}.  The average correct-sign fraction
is estimated with simulation studies and found to be 72\% and 75\% in events with 
exactly one and at least two $b$-tagged jets, respectively.
The most probable combination out of all the possible jet permutations is
chosen. Permutations with non-$b$-tagged jets assigned as $b$-jets and
vice versa have a reduced weight due to the tagging probability in the likelihood.
Finally, the lepton charge  $Q_\ell$ is used to determine if the reconstructed 
semileptonically-decaying quark is 
a top quark ($Q_\ell>0$) or an anti-top quark ($Q_\ell<0$).
The distributions of reconstructed quantities, \mtt{}, \pttt{} and \betatt{} 
are shown in Fig.~\ref{fig:datamcreco2012}, with the binnings that are used in the differential 
measurements. 

\subsection{Unfolding}
\label{sec:unfolding}

The reconstructed \dy{} distributions are distorted by acceptance 
and detector resolution effects. 
An unfolding procedure is used to estimate the 
true \dy{} spectrum, as defined by the $t$ and $\bar{t}$ after radiation and before decay in Monte Carlo events,
from the one measured in data. 
The observed spectrum is unfolded using
the fully Bayesian unfolding (FBU) technique~\cite{Fbu2012arXiv1201.4612C}. 

The FBU method consists of the strict
application of Bayesian inference to the problem of unfolding. This
application can be stated in the following terms: given an observed
spectrum $\Data$ with $N_{\mathrm r}$ reconstructed bins, 
and a response matrix
$\TrasfMatrix$ with $N_{\mathrm r} \times N_{\mathrm t}$ bins giving the detector
response to a true spectrum with $N_{\mathrm t}$  bins, the posterior
probability density of the true spectrum $\Truth{}$ (with $N_{\mathrm t}$ bins) follows
the probability density
\begin{equation}
\conditionalProb{\Truth{}}{\Data{}}
\propto{}
\conditionalLhood{\Data{}}{\Truth{}}
\cdot{}
\pi{}\left(\Truth{}\right),
\end{equation}
where  \conditionalLhood{\Data{}}{\Truth{}} is the
likelihood function of \Data{} given \Truth{} and \TrasfMatrix{}, 
and \prior{} is the prior probability density for 
\Truth{}.
While the response matrix is estimated from the simulated
sample of \ttbar{} events, a uniform prior probability density in all bins 
is chosen as \prior{}, such that equal probabilities to all \Truth{} spectra within
a wide range 
are assigned.
The unfolded asymmetry \AC{} is computed from \conditionalProb{\Truth{}}{\Data{}} as
\begin{equation}
\conditionalProb{\AC{}}{\Data{}} = \int{ \delta(\AC{} - \AC{}(\Truth{})) \conditionalProb{\Truth{}}{\Data{}} \mathrm{d} \Truth{}}.
\end{equation}

The treatment of systematic uncertainties is consistently included in the
Bayesian inference approach by extending the likelihood
\conditionalLhood{\Data{}}{\Truth{}} with nuisance parameter terms.
The marginal likelihood is defined as
\begin{equation}
\conditionalLhood{\Data{}}{\Truth{}}=
\int
\conditionalLhood{\Data{}}{\Truth{},\thetavec{}}
\cdot{}\mathcal{N}(\thetavec{})~\mathrm{d}\thetavec{},
\end{equation}
where \thetavec{} are the nuisance parameters, and
$\mathcal{N}(\thetavec{})$ their prior probability densities, which are
assumed to be Normal distributions with mean $\mu=0$ and 
standard deviation $\sigma=1$.
A nuisance parameter is associated with each of the uncertainty sources (as explained below).

The marginalisation approach provides a natural framework to treat
simultaneously the unfolding and background estimation using
multiple data regions. Given the distributions $\Data{}_i$ measured in $N_{\textrm{ch}}$
independent channels, the likelihood is extended to the product of likelihoods of each channel, 
so that
\begin{equation}
\conditionalLhood{\{\Data{}_1\cdots{}\Data{}_{N_{\mathrm ch}}\}}{\Truth{}}=
\int
\prod_{i=1}^{N_{\mathrm ch}}\conditionalLhood{\Data{}_i}{\Truth{},\thetavec{}}
\cdot{} \mathcal{N}(\thetavec{})
~\mathrm{d}\thetavec{},
\label{eq:xtllh}
\end{equation}
where the nuisance parameters are common to all analysis channels.

\subsection{Systematic uncertainties}
\label{sec:systematics}
Several sources of systematic uncertainty are considered, which can affect the normalisation of signal 
and background and/or the shape of the relevant distributions
Individual sources of systematic uncertainty are considered to be uncorrelated. Correlations of a given 
systematic uncertainty with others are maintained across signal and background processes and channels.  
The following sections describe each of the systematic uncertainties considered in the analysis. 
Experimental uncertainties and background modelling uncertainties (Sects.~\ref{sec:syst_expunc} and~\ref{sec:syst_bkgmodeling})
are marginalised during the unfolding procedure, while signal modelling uncertainties, uncertainties due to Monte Carlo sample size,
PDF uncertainties and unfolding response uncertainties (Sects.~\ref{sec:syst_sigmodeling} and~\ref{sec:syst_others}) 
are added in quadrature to the unfolded uncertainty.

\subsubsection{Experimental uncertainties}
\label{sec:syst_expunc}
\textbf{Jet energy scale and resolution:}
The jet energy scale (JES) and its uncertainty have been derived by combining
information from test-beam data, LHC collision data and
simulation~\cite{Aad:2014bia}.  The jet energy scale uncertainty is split into 22
uncorrelated components which can have different jet $\pt$ and $\eta$
dependencies and are treated independently in this analysis. 
The jet energy resolution (JER) has been determined as a function of 
jet $\pt$ and rapidity using dijet events from data and simulation. The JER in data and in simulation 
are found to agree
within 10\%, and the corresponding uncertainty is assessed by smearing the jet
$\pt$ in the simulation.
The JES and JER uncertainties represent the leading sources of uncertainty
associated with reconstructed objects in this analysis.

\textbf{Heavy- and light-flavour tagging:}
The efficiencies to tag jets from $b$-quarks, $c$-quarks, and light quarks 
are measured in data as a function of $\pt$ (and $\eta$ for light-quark jets), and these 
efficiencies are used to adjust the simulation to match data. 
The uncertainties in the calibration are propagated through this analysis and represent 
a minor source of uncertainty.

\textbf{Jet reconstruction and identification:}
The uncertainty associated with the jet reconstruction efficiency
is assessed by randomly removing 0.2\% of the jets with $\pt$ below $30~\GeV{}$, 
to match the measured jet inefficiency in data for this $\pt$ range~\cite{Aad:2014bia}.
The uncertainty on the efficiency that each jet satisfies 
the JVF requirement is estimated by changing the JVF cut value
from its nominal value by $\pm0.1$, and repeating the analysis using the modified cut value.
Both uncertainties have a negligible impact on the measurement.

\textbf{Leptons:}
Uncertainties associated with leptons affect the reconstruction,
identification and trigger efficiencies, as well as the lepton momentum scale and
resolution. They are estimated from $Z\to \ell^+\ell^-$ ($\ell=e,\mu$), 
$J/\psi \to \ell^+\ell^-$ and $W\to e\nu$ processes using techniques 
described in Refs.~\cite{Aad:2014fxa,Aad:2011mk,Aad:2014zya}.
The combined effect of all these uncertainties results in an overall normalisation 
uncertainty on the signal and background of approximately 1.5\%.
Charge misidentification is not considered as it is small~\cite{Aad:2011mk}
and has a negligible impact on the measurement.

\textbf{Missing transverse momentum:}
The $\met$ reconstruction is affected by uncertainties associated with leptons, jet 
energy scales and resolutions which are propagated to the $\met$ calculation. 
Additional small uncertainties 
associated with the modelling of the underlying event, in particular its impact on
the $\pt$ scale and resolution of unclustered energy, are also taken into account.
All uncertainties associated with the $\met$ have a negligible effect.

\textbf{Luminosity:}
The uncertainty on the integrated luminosity is 2.8\%, affecting the overall normalisation of
all processes estimated from MC simulation. It is derived following the methodology 
detailed in Ref.~\cite{Aad:2013ucp}. The impact of this uncertainty is negligible in this measurement.

\subsubsection{Background modelling}
\label{sec:syst_bkgmodeling}

\textbf{\wjets{}:}
The predictions of normalisation and flavour composition of the $W$+jets background are 
affected by large uncertainties, but the in situ data-driven technique described in 
Sect.~\ref{sec:wzjets} reduces these to a negligible level. 
All sources of uncertainty other than normalisation are propagated to the \wjets{} estimation.

\textbf{\zjets{}:}
Uncertainties affecting the modelling of the $Z$+jets background include a 5\%  
normalisation uncertainty from the theoretical NNLO cross section~\cite{Melnikov:2006kv}, 
as well as an additional 24\% normalisation uncertainty added in quadrature for each additional 
inclusive jet-multiplicity bin, based on a comparison among different algorithms 
for merging LO matrix elements and parton showers~\cite{Alwall:2007fs}.
The normalisation uncertainties for $Z$+jets are described by three uncorrelated 
nuisance parameters corresponding to the 
three $b$-tag multiplicities considered in the analysis.

\textbf{Multijet background:}
Uncertainties on the multijet background estimated via the Matrix Method receive 
contributions from the size of the data sample as well as from the uncertainty on $\epsf$, estimated in different control regions. A normalisation uncertainty of 50\% due 
to all these effects is assigned independently to the electron and muon channels and to each $b$-tag multiplicity, 
leading to a total of six uncorrelated uncertainties.

\textbf{Other physics backgrounds:}
Uncertainties affecting the normalisation of the single-top-quark background include a 
+5\%/$-4\%$ uncertainty on the total cross section estimated as a weighted average 
of the theoretical uncertainties on $t$-, $Wt$- and $s$-channel 
production~\cite{Kidonakis:2011wy,Kidonakis:2010ux,Kidonakis:2010tc}. Including an additional uncertainty in 
quadrature of 24\% per additional jet has a negligible impact on the measurement.
Uncertainties on the diboson background normalisation include 5\% from the
NLO theoretical cross sections~\cite{Campbell:1999ah} added in quadrature 
to an uncertainty of 24\% due to the extrapolation to the high jet-multiplicity region, 
following the procedure described for $Z$+jets.

\subsubsection{Signal modelling}
\label{sec:syst_sigmodeling}
In order to investigate the impact of uncertainties on the \ttbar{} signal modelling, additional samples generated
with {\textsc Powheg-Box} interfaced to \herwig{}, \mcnlo{} interfaced to \herwig{} and {\textsc AcerMC} interfaced to {\textsc Pythia} 
are considered (see Sect.~\ref{sec:ttbarmod} for more details). Different predictions and response matrices built with those 
\ttbar{} samples are used to repeat the full analysis procedure isolating one effect at the time. For each case, the intrinsic 
asymmetry and the unfolded asymmetry are measured. The intrinsic asymmetry is the asymmetry generated in each Monte 
Carlo sample before the simulation of the detector response. Double differencees between the intrinsic (int) asymmetry and 
the unfolded (unf) values of the nominal (nom) and the alternative (alt) sample are considered as uncertainties to account 
for the different \AC{} predictions of the different samples, $(\AC{}^{\mathrm{int, nom}} - \AC{}^{\mathrm{int, alt}}) - (\AC{}^{\mathrm{unf, nom}} - \AC{}^\mathrm{{unf, alt}})$. This is referred to as the double difference.

\textbf{NLO generator:}
The uncertainty 
associated with the choice of NLO generator is estimated from the double difference of the 
parton-level \AC{} and unfolded \AC{} comparing {\textsc Powheg-Box} interfaced to \herwig{} (nom) and 
\mcnlo{} interfaced to \herwig{} (alt). 

\textbf{Fragmentation model:}
The uncertainty associated with the fragmentation model is 
estimated from the double difference of the parton-level \AC{} and 
unfolded \AC{} comparing {\textsc Powheg-Box} interfaced to \pythia{} (nom) and {\textsc Powheg-Box} interfaced to \herwig{} (alt).

\textbf{Initial- and final-state radiation (ISR/FSR):}
The uncertainty associated with the ISR/FSR modelling is estimated using the {\textsc AcerMC} generator
where the parameters of the generation 
were varied to be compatible with the results of a measurement of $t\bar{t}$
production with a veto on additional central jet activity~\cite{ATLAS:2012al}. 
Two variations producing more and less ISR/FSR are considered.
The uncertainty is estimated from half of the double difference 
of the parton-level \AC{} and unfolded \AC{} comparing {\textsc Powheg-Box} (nom) and 
{\textsc AcerMC} (alt)  interfaced to \pythia{} producing more and less ISR/FSR.

\subsubsection{Others}
\label{sec:syst_others}
\textbf{Monte Carlo sample size:}
 To assess the effect on the measurement of the limited number of Monte Carlo events, 
 an ensemble of 1\,000 response matrices, each of them
 fluctuated according to the raw number of simulated events, is produced. 
 Unfolding is repeated with the same pseudo-dataset for each fluctuated response matrix. 
 The uncertainty is estimated as the standard deviation of the ensemble of the 1\,000 \AC{} values obtained.
The estimated systematic uncertainty associated with limited number of Monte Carlo events is about ten times 
smaller than the data statistical uncertainty; this is consistent with the size of the available Monte Carlo sample.

\textbf{PDF uncertainties:}
The choice of PDF in simulation 
has a significant impact on the charge asymmetry of the simulated \wjets{} background.
Since this asymmetry is exploited to calibrate the \wjets{} prediction,
the related uncertainty has to be estimated.
The uncertainty on the PDFs is evaluated 
using three different PDF sets: CT10~\cite{Lai:2010vv}, MSTW 2008~\cite{mstw} and NNPDF2.1~\cite{nnpdf}.
For each set, the PDFs are varied based on the uncertainties along each of the PDF eigenvectors.
Each variation is applied by reweighting the \wjets{} sample event-by-event.
The \AC{} measurements are repeated for each varied \wjets{} template and the uncertainty
is estimated as half of the largest difference between 
any variation of CT10 and MSTW 2008, and the $\pm1\sigma$ variations for NNPDF2.1. The resulting 
uncertainties are small, but non-negligible. The impact of uncertainties related to PDFs are found to be negligible in \ttbar{} modelling.

\textbf{Unfolding response:}
The response of the unfolding procedure, i.e. any non-linearity or bias, is determined using a set of six
pseudo-datasets, each of them being composed of the default \ttbar{} signal 
reweighted to simulate an asymmetry and the default MC simulation predictions.
The injected \AC{} value ranges between $-0.2$ and 0.2 
depending on the differential variable and bin.
The six reweighted pseudo-datasets are unfolded using the default response matrix 
and the uncertainty associated with the unfolding response is calculated as: 
$A_\mathrm{C}^{\mathrm{meas}} - (A_\mathrm{C}^{\mathrm{meas}}- b)/a$, with $a$ 
and $b$ the slope and offset of a linear fit of the generator-level (intrinsic) \AC{} versus unfolded \AC{} 
of the six reweighted pseudo-datasets previously defined and $A_\mathrm{C}^{\mathrm{meas}}$ 
the measured value in data.

\begin{figure}[!ht]\centering
\includegraphics[width=0.68\textwidth]{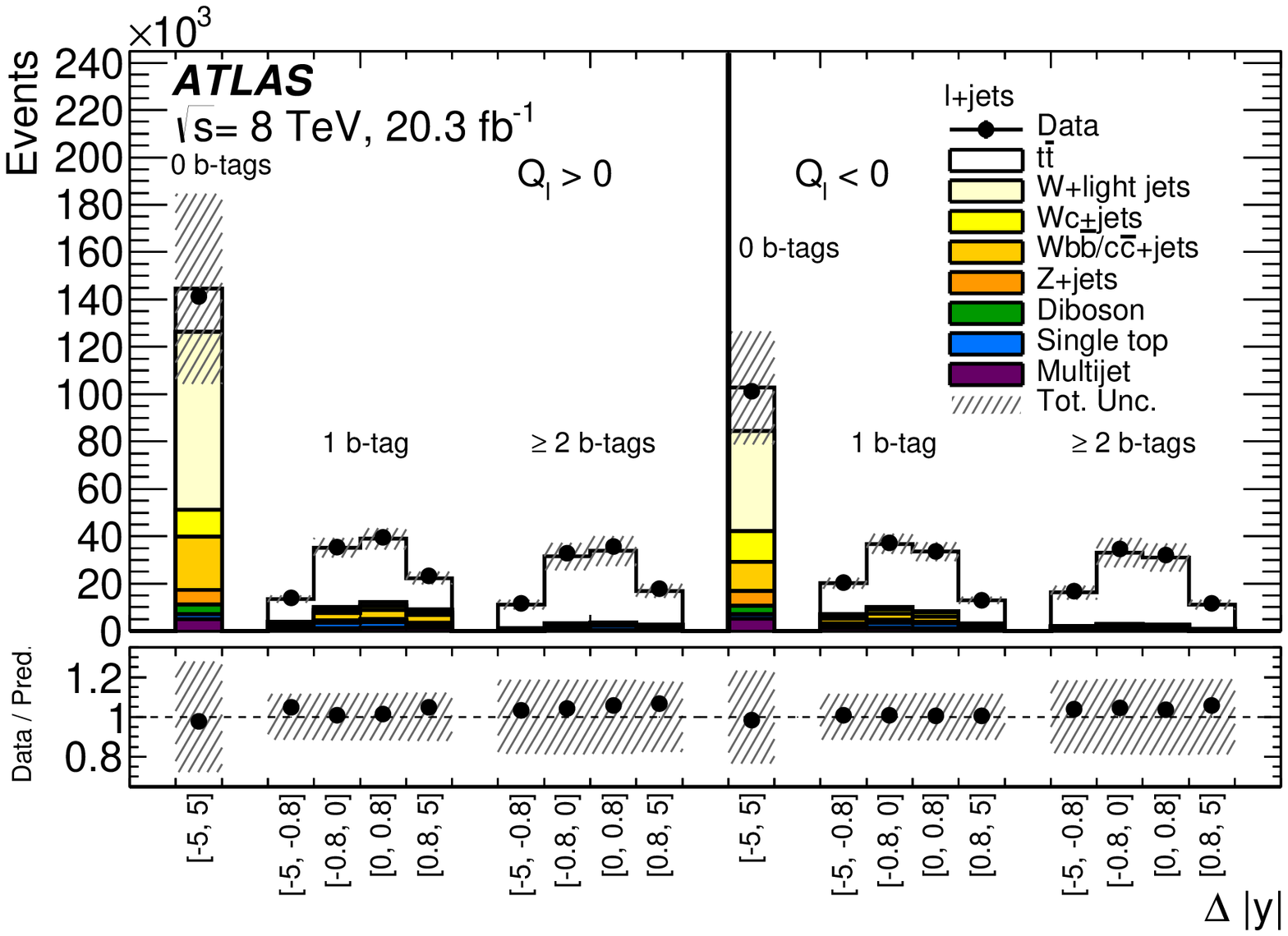}
\includegraphics[width=0.68\textwidth]{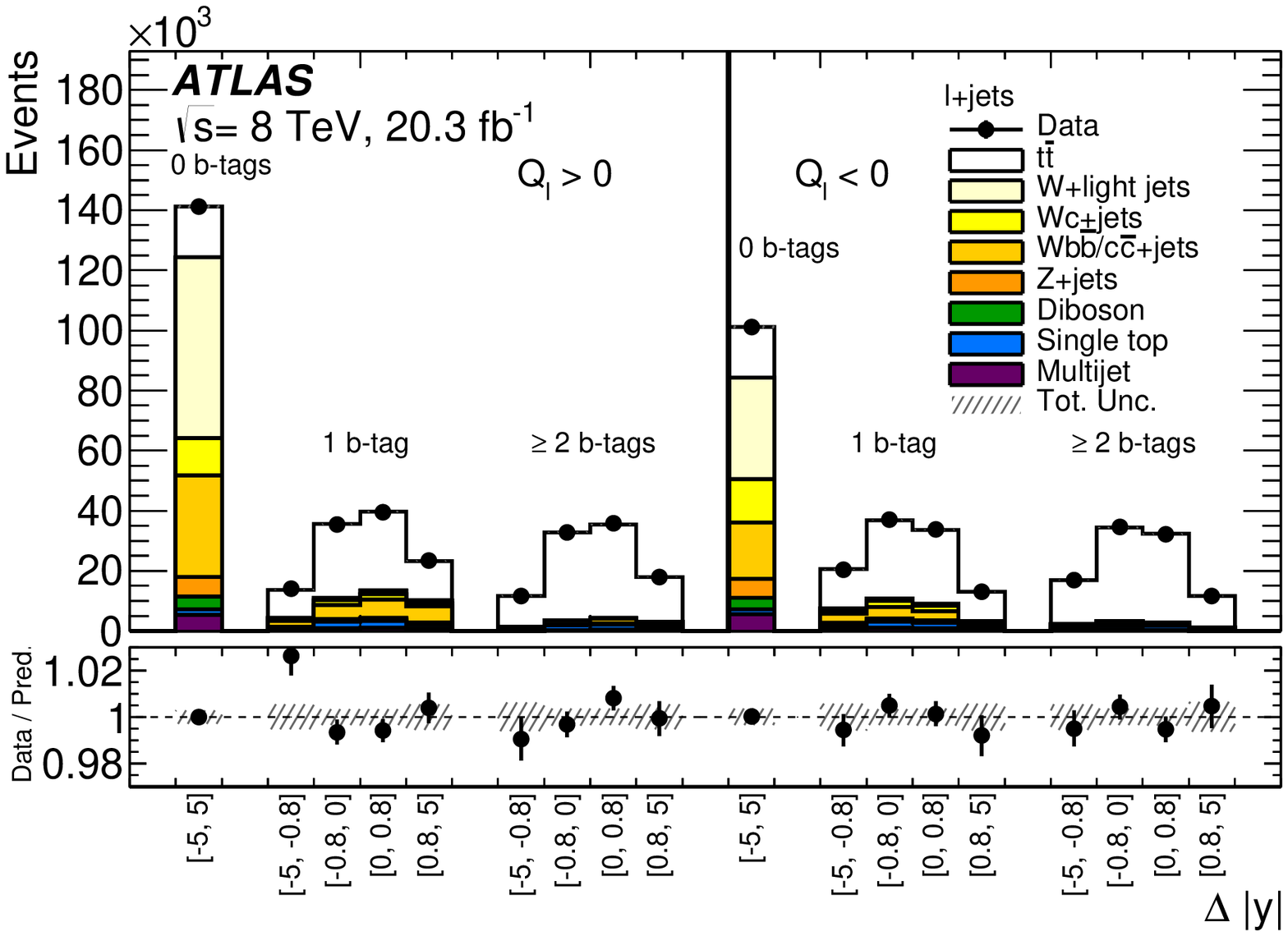}

\caption{Comparison between prediction and data for the 18 bins used in the inclusive \AC{}
measurement before (top) and after (bottom) the simultaneous unfolding procedure and $W$+jets in situ background
calibration, including only uncertainties that are marginalized. The \dy{} distribution in four bins is considered for the \ttbar{}-enriched event samples with exactly one
and at least two $b$-jets; a single bin is considered for the background-enriched sample with zero $b$-jets.
After the calibration, the background components are 
scaled to the measured values for the nuisance
parameters, and the prediction for \ttbar{} events in each bin is estimated by folding the measured
parton-level parameters through the response matrix. 
The bottom part of each plot shows the ratio of the data to the predicted value together with
combined statistical and systematic uncertainties.}
\label{fig:datamcsix}
\end{figure}

\subsection{Measurement}
\label{sec:measurement}
A fit is performed which maximises the extended likelihood of Eq.~(\ref{eq:xtllh}). In this fit, 
the events are further separated based on the sign of the lepton charge $Q_\ell$.
The measurements are then performed using a combination of six channels based on
the lepton charge ($Q_\ell>0$ and $Q_\ell<0$) and the $b$-jet multiplicity
(zero $b$-jets, one $b$-jet, at least two $b$-jets). 
The \dy{} distribution is split into four bins in all 
the channels except the zero $b$-jets channel, as no extra information for \AC{} is expected.
Four bins in \dy{} are considered in each differential bin of all differential measurements.

The \wjets{} in situ calibration procedure consists of fitting the calibration factors 
$K_{\bbbar{}/\ccbar{}}$, $K_c$ and $K_{\text{light}}$ for scaling the flavor components 
of the \wjets{} background with different charge asymmetries, assuming uniform prior probabilities \pi{}
during the posterior probability estimation defined in Eq.~(\ref{eq:wjets}).
The $b$-jet multiplicity provides
information about the heavy- and light-flavour composition of the \wjets{} background, 
while the lepton charge asymmetry is used to determine the normalisation of each component. 
Figure~\ref{fig:datamcsix} shows the different \wjets{} contributions for the different 
$b$-jet multiplicities and lepton charges.
In addition to the expected number of \ttbar{} events for each bin in \Truth{}, the
\wjets{} calibration factors are free parameters in the likelihood.
The posterior probability density is thus
\begin{equation}
\label{eq:wjets}
\begin{split}
\conditionalProb{\Truth{}}{\{\Data{}_1\cdots{}\Data{}_{N_{\mathrm ch}}\}}=&
\int
\prod_{i=1}^{N_{\mathrm ch}}\conditionalLhood{\Data{}_i}{\Reco{}_i(\Truth{};\thetavec{}_{\mathrm s}),\Bckg{}_i(K_{\bbbar{}/\ccbar{}},K_c,K_{\text{light}};\thetavec{}_{\mathrm s},\thetavec{}_{\mathrm b})}\\
& ~\mathcal{N}(\thetavec{}_{\mathrm s})
~\mathcal{N}(\thetavec{}_{\mathrm b})
~\pi{}(\Truth{})
~\pi{}(K_{\bbbar{}/\ccbar{}})
~\pi{}(K_c)
~\pi{}(K_{\text{light}})
~\mathrm{d}\thetavec{}_{\mathrm s}
~\mathrm{d}\thetavec{}_{\mathrm b},\\
\end{split}
\end{equation}
where
$\Bckg{}=\Bckg{}(K_{\bbbar{}/\ccbar{}},K_c,K_{\text{light}};\thetavec_{\mathrm s},\thetavec{}_{\mathrm b})$
is the total background prediction,  the probability densities $\pi{}$ are uniform priors 
and \Reco{} is the reconstructed signal prediction.
Two categories of nuisance parameters are considered: 
the normalisation of the background processes
($\thetavec{}_{\mathrm b}$), and the uncertainties associated with the object
identification, reconstruction and calibration ($\thetavec{}_{\mathrm s}$).
While the first ones only affect the background predictions, the
latter, referred to as object systematic uncertainties, affect both the
reconstructed distribution for \ttbar{} signal and the total background prediction.
The \wjets{} calibration factors are found to be $K_{\bbbar{}/\ccbar{}}=1.50\pm0.11$, 
$K_c = 1.07\pm0.27$ and $K_{\text{light}} = 0.80\pm0.04$, 
where the uncertainties include both the statistical and systematic components.

The final numbers of expected and observed data events after the full event selection, marginalisation of nuisance parameters and 
$W$+jets in situ  calibration  are listed in Table~\ref{tab:evtnumbers}, while Fig.~\ref{fig:datamcsix} 
shows the good level of agreement between the data and expectation
 before and after marginalisation for the six channels. In both cases, the uncertainties that are marginalized are shown.  Since these 
uncertainties are correlated for the background and signal components, the total combined marginalized uncertainty is smaller than the 
sum of the constituent parts.

\begin{table}[!ht]
  \begin{center}
      \caption{Observed number of data events compared to the expected number
  of signal events and different background contributions for different $b$-tagging multiplicities
   in the combined \mujets{} and \ejets{} channels. These yields are
shown after marginalisation of the nuisance parameters and the in situ calibration of the  $W$+jets background, and the marginalized uncertainties are shown. The marginalized 
uncertainties for each background and signal component are correlated, and the correlation is taken into account in their combination.
  }
  \label{tab:evtnumbers}
  {\footnotesize
  \begin{tabular}{ l   rrr  rrr  rrr }

  \toprule
      Channel & \multicolumn{3}{c}{\ljets{} 0-tag} & \multicolumn{3}{c}{\ljets{} 1-tag} & \multicolumn{3}{c}{\ljets{} 2-tag} \\
      \midrule
      Single top       & 3400  &$\!\!\!\pm\!\!\!$& 400  & 12100  &$\!\!\!\pm\!\!\!$& 1300  & 8700  &$\!\!\!\pm\!\!\!$& 900   \\
      $W$+jets        & 173000  &$\!\!\!\pm\!\!\!$& 9000  & 45000 &$\!\!\!\pm\!\!\!$& 4000  & 8600 &$\!\!\!\pm\!\!\!$& 700  \\
      $Z$+jets         & 13000   &$\!\!\!\pm\!\!\!$& 6000  & 3900  &$\!\!\!\pm\!\!\!$& 2000  & 1900 &$\!\!\!\pm\!\!\!$& 900   \\
      Diboson          & 8000  &$\!\!\!\pm\!\!\!$& 4000 & 2000 &$\!\!\!\pm\!\!\!$& 900 & 400  &$\!\!\!\pm\!\!\!$& 200 \\
      Multijets         & 10800  &$\!\!\!\pm\!\!\!$& 3500  & 6300  &$\!\!\!\pm\!\!\!$& 2000  & 2200   &$\!\!\!\pm\!\!\!$& 700  \\
      \midrule
      Total background & 208500  &$\!\!\!\pm\!\!\!$& 1300  & 69600 &$\!\!\!\pm\!\!\!$& 2600 & 21800  &$\!\!\!\pm\!\!\!$& 1300  \\
      $t\bar{t}$        & 33900 &$\!\!\!\pm\!\!\!$& 1200  & 146900 &$\!\!\!\pm\!\!\!$& 2700  & 171600 &$\!\!\!\pm\!\!\!$& 1500   \\
      \midrule
      Total expected   & 242400 &$\!\!\!\pm\!\!\!$& 600  & 216500 &$\!\!\!\pm\!\!\!$& 500 & 193400 &$\!\!\!\pm\!\!\!$& 400  \\
      \midrule
      Observed         & \multicolumn{3}{c}{242420}     & \multicolumn{3}{c}{216465}     & \multicolumn{3}{c}{193418}     \\
      \bottomrule
          \end{tabular}
  }
  \end{center}
\end{table}

%% file: results.tex
\section{Results}
\label{sec:result}

\subsection{Inclusive measurement}
\label{sec:incmeas}
The inclusive \ttbar\ production charge asymmetry is measured to be 
$$\AC{}= 0.009\pm0.005~(\textrm{stat.}+\textrm{syst.})\textrm{,}$$ compatible with the SM prediction,
$\AC{}=0.0111\pm0.0004$~\cite{Bern:2012}.

Since the background estimation is part of the Bayesian inference
procedure described in Sect.~\ref{sec:unfolding}, it is not possible to
study the impact of systematic uncertainties by repeating unfolding on data with
varied templates, without using marginalisation. Instead, the expected
impact of systematic uncertainties is studied with pseudo-data distributions
corresponding to the sum of the background and signal predictions.
For each source of uncertainty, the $\pm{}1\sigma$ variations of the
predictions are used to build the pseudo--data, and the unfolding
procedure is repeated. The baseline background templates and response
matrices, as in the actual measurements, are used.
Table~\ref{tab:8tevsystematics} shows the average asymmetry variation
$\delta{\AC{}}$ computed, for each source of uncertainty, as
$|\AC{}(+1\sigma)-\AC{}(-1\sigma)|/2$, but only the uncertainties having a variation 
above $10\%$ of the statistical uncertainty are reported in the table.
The total uncertainty associated with the marginalised systematic uncertainties is estimated by subtracting 
in quadrature the statistical term from the total marginalised uncertainty. 
It yields $0.002$ (category (a) in Table~\ref{tab:8tevsystematics}). 
The total, non-marginalised uncertainty associated with systematic uncertainties is estimated by summing in quadrature sources from category (b) in Table~\ref{tab:8tevsystematics}.

\begin{table}[!ht]\centering
\caption{
Impact of individual sources of uncertainty on the inclusive
\AC{} measurement. All uncertainties described in Sect.~\ref{sec:systematics}
are considered, but only the ones having a variation
 above $10\%$ of the statistical uncertainty are reported in the table.
 Systematic uncertainties in group (a) are marginalised while systematic uncertainties in group (b) are
added in quadrature to the marginalised posterior.}
\label{tab:8tevsystematics}
\begin{tabular}{l l c}
\toprule
& Source of systematic uncertainty & $\delta{\AC{}}$ \\
\midrule
(a)& Jet energy scale and resolution & $0.0016$ \\
& Multijet background normalisation & $0.0005$\\
\midrule
(b)& Initial-/final-state radiation & $0.0009$\\
& Monte Carlo sample size & $0.0010$\\
& PDF &$0.0007$\\
\midrule
& Statistical uncertainty & $0.0044$ \\
\midrule
& Total uncertainty & $0.0049$ \\
\bottomrule
\end{tabular}
\end{table}

The precision of the measurement is limited by
the statistical uncertainty, and the main sources of systematic
uncertainty are the signal modelling and the uncertainties with a large impact on the size of the
\wjets{} background, such as the uncertainty on the jet energy scale
and resolution.

\subsection{Differential measurements}
\label{sec:diffmeas}

\begin{figure}[!htb]\centering
\includegraphics[width=0.49\textwidth]{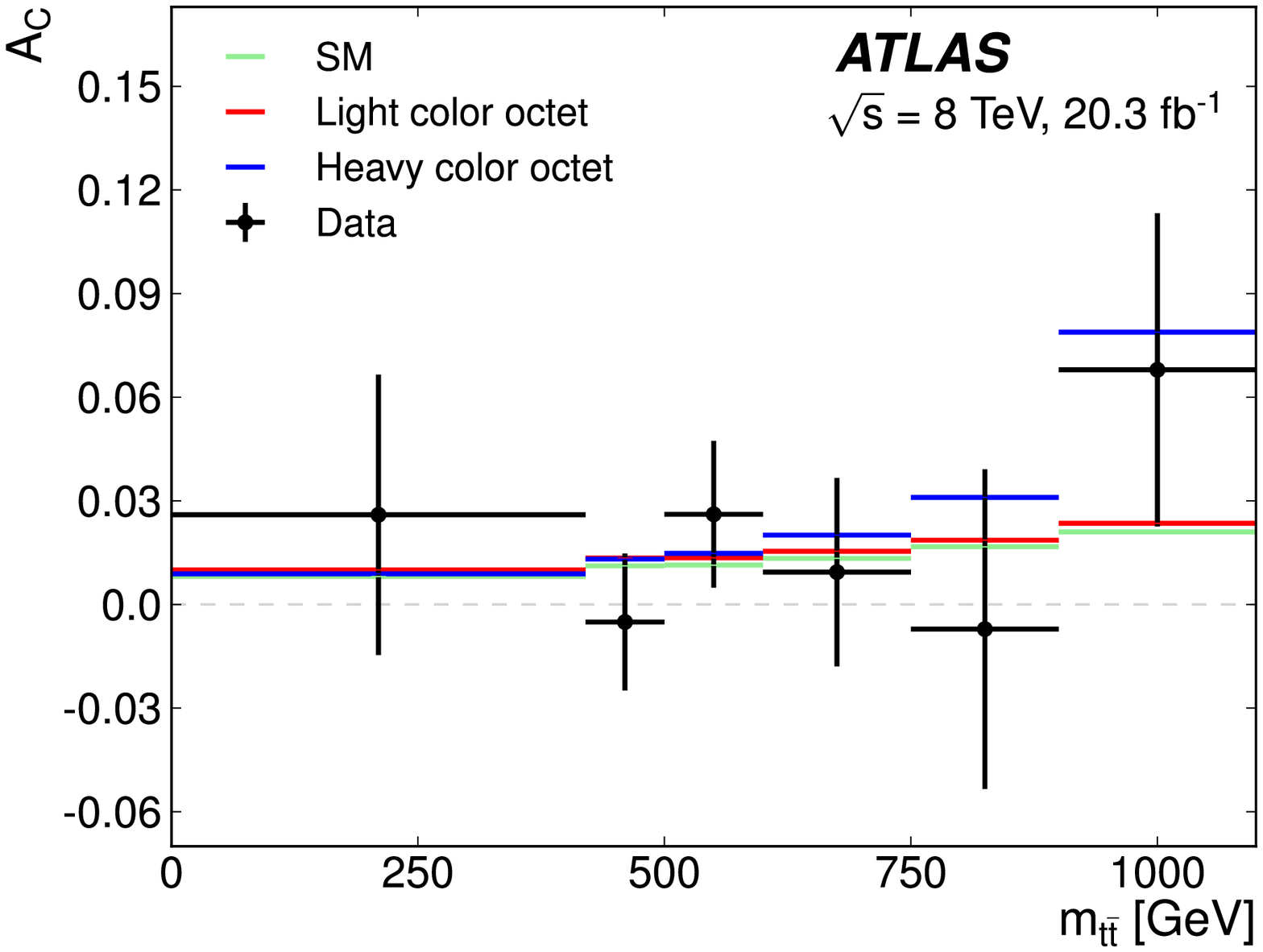}
\includegraphics[width=0.495\textwidth]{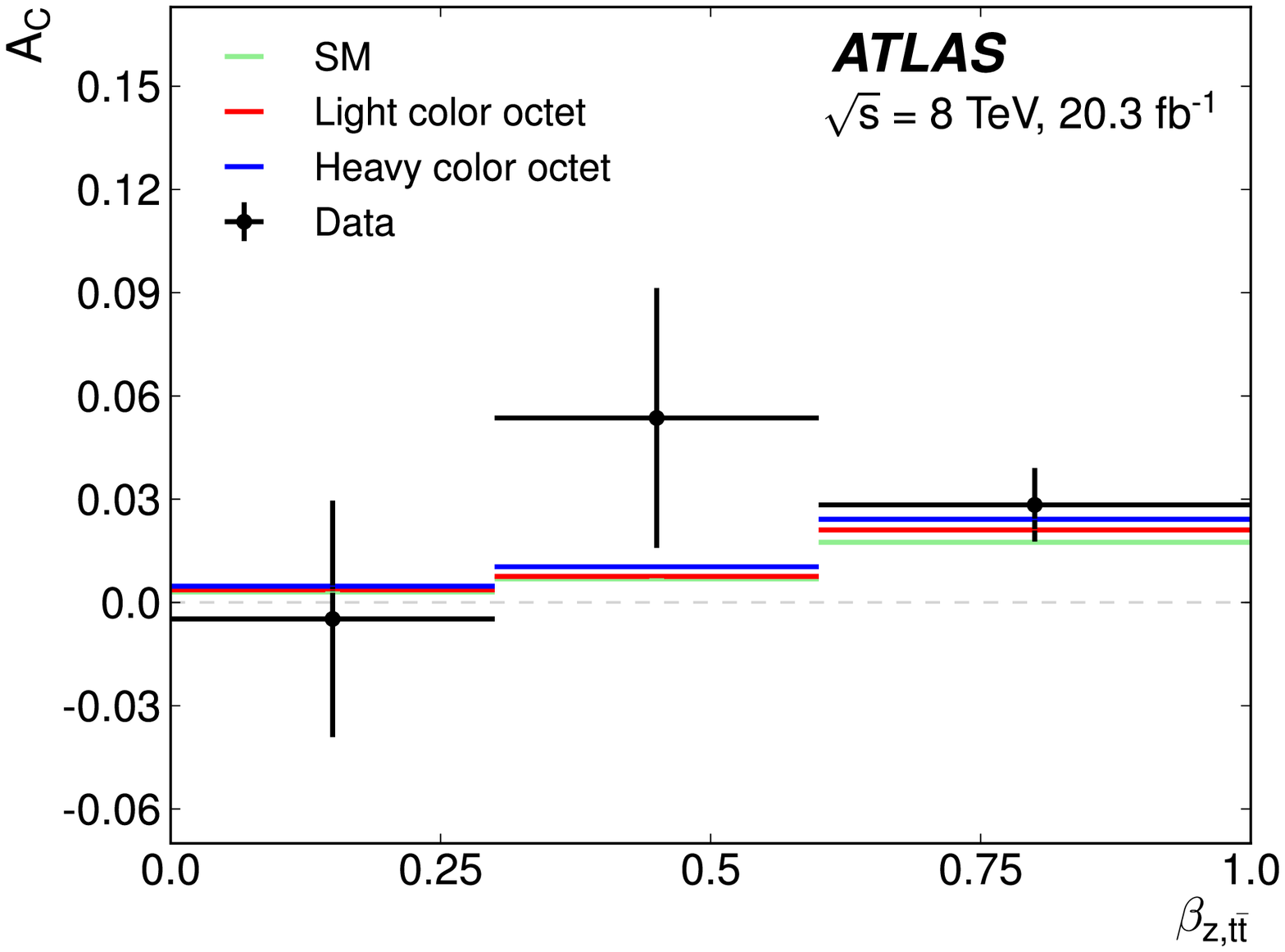}
\includegraphics[width=0.49\textwidth]{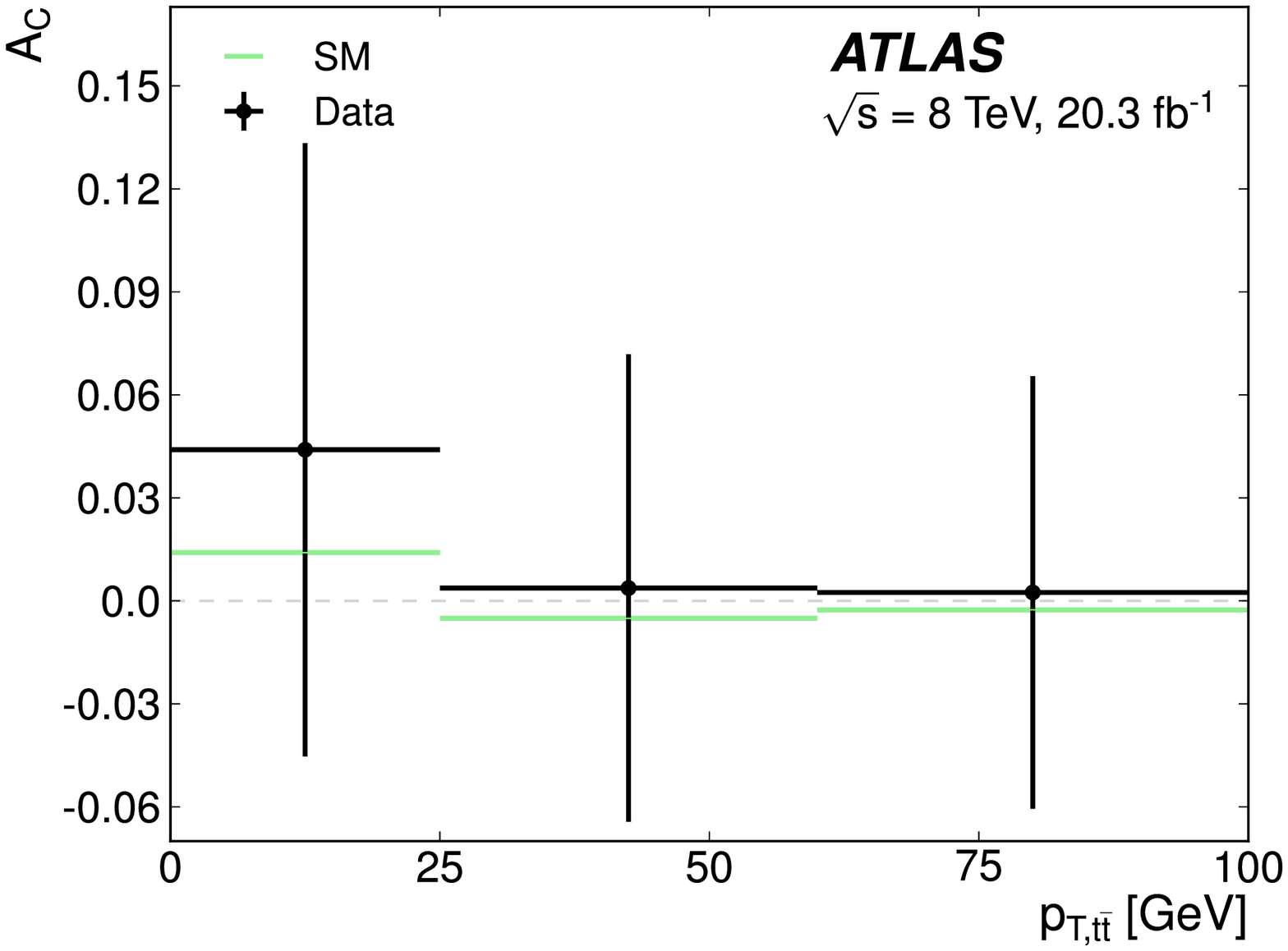}
\caption{Measured \AC{} values as a function of bin-averaged \mtt{}, \betatt{} and \pttt{},
compared with predictions for SM~\cite{Bern:2012}
and for right-handed
colour octets with masses below the \ttbar{} threshold and beyond
the kinematic reach of current LHC searches~\cite{Aguilar-Saavedra:2014nja}.
The BSM predictions are shown only for the two top plots.
The bins are the same as the ones reported in
Table~\ref{tab:systdiff} and Table~\ref{tab:results_diff}.}
\label{fig:8tevacvsmtt}
\end{figure}

The \AC{} differential spectra are compared in 
Fig.~\ref{fig:8tevacvsmtt} with the theoretical SM predictions, 
as well as with BSM predictions for right-handed colour octets with low and high
masses~\cite{Aguilar-Saavedra:2014nja}. The BSM predictions are not shown in the 
measurement as a function of \pttt{} as they are LO $2\to2$ calculations. 
The results are compatible with the SM, and 
it is not possible to distinguish between the SM and BSM models at this level of precision.
The BSM models are
tuned to be compatible with the Tevatron asymmetry measurements and the
\AC{} measurements at $\sqrt{s}=7$ TeV.

\begin{table}[!htb]\centering
  \caption{Impact of individual sources of uncertainty on the measurement of $\AC{}$ in bins of $\mtt{}$, $\betatt{}$ and $\pttt{}$.
  All uncertainties described in Sect.~\ref{sec:systematics} are considered, but only the ones having at least one bin with a variation 
  above $10\%$ of the statistical uncertainty are reported in the table; the others are quoted 
   as ``$-$''. Systematic uncertainties in group (a) are marginalised while systematic uncertainties in group (b) are
added in quadrature to the marginalised posterior. }
  \label{tab:systdiff}

   \vskip 0.5cm

  {\footnotesize
    \begin{tabular}{l l c c c c c c}
      \toprule
        & \multicolumn{7}{c}{$\delta A_{\mathrm C}$ in $\mtt{}$  [\GeV{}]  } \\
        & Source of systematic uncertainty  &  $0$--$420$  &  $420$--$500$  &  $500$--$600$  &  $600$--$750$  &  $750$--$900$  &  $>900$ \\
      \midrule 
	(a)& Jet energy scale and resolution      	& 0.010	& 0.007	& 0.007	& 0.009	& 0.013	& 0.009 	\\
	& $b$-tagging/mis-tag efficiencies   	& 0.006	& 0.005	& 0.005	& 0.005	& 0.008	& 0.005	\\
	& Missing transverse momentum 		& $-$	& $-$	& 0.003	& 0.002	& $-$	& $-$	\\
	& Lepton reconstruction/identification 	& 0.004	& $-$	& $-$	& $-$	& $-$	& $-$	\\
	& Other backgrounds normalisation		& 0.009	& 0.006	& $-$	& 0.002	& $-$	& $-$	\\
\midrule
	(b)& Signal modelling             			& 0.030	&  0.005	&  0.004	& 0.009	& $-$	& 0.007    \\ 
	& Parton shower/hadronisation 		      	& $-$	&  0.005	&  $-$	& $-$	& 0.010	& 0.011    \\ 
	& Initial-/final-state radiation			& 0.006	&  0.002	& 0.004    & 0.004	& 0.004	& 0.011    \\ 
	& Monte Carlo sample size      			& 0.006    	&  0.004 	& 0.004	& 0.005	& 0.010	& 0.009    \\ 
	& PDF                         				& 0.004	&  0.002	& 0.002	& 0.004	& 0.005     & 0.007    \\
     \midrule
	& Statistical uncertainty     			& 0.025    	&  0.017    & 0.018	& 0.023	& 0.042     & 0.037 \\
      \midrule
	& Total              				                & 0.041	&  0.020    & 0.021	& 0.027    & 0.046    & 0.045 \\
      \bottomrule
    \end{tabular}
  }
   \vskip 0.5cm

  {\footnotesize
    \begin{tabular}{l l c c c}
      \toprule
      & \multicolumn{4}{c}{$\delta A_{\mathrm C}$ in $\betatt{}$} \\
  	& Source of systematic uncertainty &  $<0.3$  &  $0.3$--$0.6$  &  $0.6$--$1.0$  \\
      \midrule 
	(a)& Jet energy scale and resolution      	& 0.009	& 0.013	& 0.003 \\
	& $b$-tagging/mis-tag efficiencies   	& 0.003	& 0.003	& 0.001 \\
	& Multijet background normalisation    	& 0.003	& $-$	& $-$ \\
\midrule
	(b)& Signal modelling             			& 0.025	& 0.027	& 0.002     \\
	& Parton shower/hadronisation 			& 0.009	& 0.010 	& 0.006     \\
	& Initial-/final-state radiation			& 0.006    & $-$ 	& $-$  \\

	& Monte Carlo sample size      			& 0.005	& 0.004	& 0.002     \\
	& PDF                         				& 0.004	& 0.006	& 0.002      \\
	\midrule
	& Statistical uncertainty     			& 0.018    & 0.015    & 0.008 \\
      	\midrule
	& Total                       				& 0.034    & 0.038    & 0.011 \\
      	\bottomrule
    \end{tabular}
 }
   
   \vskip 0.5cm
   
  {\footnotesize
    \begin{tabular}{l l c c c}
      \toprule
       	& \multicolumn{4}{c}{$\delta A_{\mathrm C}$ in $\pttt{}$ [\GeV{}]} \\
	& Source of systematic uncertainty &  $0$--$25$ & $25$--$60$ & $>60$ \\
      \midrule  
	(a)& Jet energy scale and resolution      	& 0.009 	& 0.009 	& 0.003 \\
	& Lepton energy scale and resolution	& 0.001	& $-$ 	& 0.003 \\
	& $b$-tagging/mis-tag efficiencies   	& 0.007 	& 0.008 	& 0.003 \\
	& Missing transverse momentum 		& 0.002 	& 0.004	& 0.002 \\
	& Multijet background normalisation    	& 0.005 	& 0.003 	& $-$ \\
	& Lepton reconstruction/identification 	& 0.005 	& 0.004 	& 0.001 \\
	& Other backgrounds normalisation		& $-$ 	& 0.003	& 0.002 \\
\midrule
	(b)& Signal modelling              			& 0.067    	& 0.017    & 0.057    \\  
	& Parton shower/hadronisation  		& 0.040	& 0.043    & 0.019    \\
	& Initial-/final-state radiation			& 0.015    & 0.017    & 0.009    \\

	& Monte Carlo sample size       			& 0.006	& 0.008    & 0.003    \\
	& PDF                          				& 0.009	& 0.009    & 0.004     \\
      \midrule
	& Statistical uncertainty      			& 0.017    & 0.028   & 0.014 \\
      \midrule
	& Total                        				& 0.089    & 0.068   & 0.063 \\
      \bottomrule
    \end{tabular}
}

\end{table}

Table~\ref{tab:systdiff} shows the average asymmetry variation
$\delta{\AC{}}$ computed for each differential measurement, for each source of uncertainty, 
as explained in Sect.~\ref{sec:incmeas}.
The precision of the differential measurements is limited by the same factors as the inclusive result.
The measurement versus \pttt{} is particularly affected by the 
parton-shower model.

The resulting charge asymmetry \AC{} is shown in Table~\ref{tab:results_diff} for the differential measurements
as a function of \mtt{} \betatt{} and \pttt{}.  The theoretical values are described
in Ref.~\cite{Bern:2012} (SM) and Ref.~\cite{Aguilar-Saavedra:2014nja} (BSM),
and they have been provided for the chosen bins.
The correlation matrices are shown in Table~\ref{table:corr_diff} for the measurements as a function of \mtt{}, \betatt{} and \pttt{}.

In regions with sensitivity to BSM (high values of \mtt{} and \betatt{}),
the uncertainty on the measurements is largely dominated by the available statistics, 
while in other regions the uncertainty on signal modeling and/or parton shower dominates.

\begin{table}[ht]
    \centering
    \caption{
  Measured charge asymmetry, \AC{}, values for the electron and muon channels combined after
unfolding as a function of the \ttbar{} invariant mass, \mtt{} (top), the \ttbar{} velocity 
along the z-axis, \betatt{} (middle), and the \ttbar{} transverse momentum, \pttt{} (bottom). 
SM and BSM predictions, for right--handed colour octets with masses below the \ttbar{} threshold (Light BSM) and beyond
the kinematic reach of current LHC searches (Heavy BSM)~\cite{Aguilar-Saavedra:2014nja}, are also reported. The quoted uncertainties include statistical and systematic components after the marginalisation.}
  \label{tab:results_diff}

  {\footnotesize
  \begin{tabular}{l c c c c c c}
  \toprule
  & \multicolumn{6}{c}{$\mtt{}$ [\GeV{}]} \\
  $\AC{}$ &  $<420$  &  $420$--$500$  &  $500$--$600$  &  $600$--$750$  &  $750$--$900$  &  $>900$ \\  
      \midrule

  Data      & 0.026 $\pm$ 0.041  & $-0.005$ $\pm$ 0.020  & 0.026 $\pm$ 0.021 &
                  0.009 $\pm$ 0.027  & $-0.007$ $\pm$ 0.046  & 0.068 $\pm$ 0.044 \\

         \midrule      

      SM        & 0.0081$^{+0.0003}_{-0.0004}$  &  0.0112 $\pm$ 0.0005          & 0.0114$^{+0.0003}_{-0.0004}$  & 
                  0.0134$^{+0.0003}_{-0.0005}$  &  0.0167$^{+0.0005}_{-0.0006}$ & 0.0210$^{+0.0003}_{-0.0002}$  \\
      \midrule      
      Light BSM & 0.0100 $\pm$ 0.0004           &  0.0134 $\pm$ 0.0006          & 0.0135$^{+0.0004}_{-0.0005}$  & 
                  0.0155$^{+0.0005}_{-0.0006}$  &  0.0186$^{+0.0007}_{-0.0008}$ & 0.0235$^{+0.0006}_{-0.0005}$ \\
      \midrule      
      Heavy BSM & 0.0089 $\pm$ 0.0004           &  0.0132 $\pm$ 0.0006          & 0.0148$^{+0.0004}_{-0.0005}$  & 
                  0.0201$^{+0.0004}_{-0.0006}$  &  0.0310$^{+0.0006}_{-0.0007}$ & 0.0788$^{+0.0007}_{-0.0006}$ \\
      \bottomrule
    \end{tabular}
  }
    \vskip 0.5cm

  {\footnotesize
    \begin{tabular}{l c c c}
  \toprule
  & \multicolumn{3}{c}{$\betatt{}$} \\
  $\AC{}$ &  $<0.3$  &  $0.3$--$0.6$  &  $0.6$--$1.0$  \\
      \midrule
            Data      & $-0.005$ $\pm$ 0.034  & 0.054 $\pm$ 0.038  & 0.028 $\pm$ 0.011 \\	
      \midrule      
      SM        &  0.0031 $\pm$ 0.0003  & 0.0068 $^{+0.0002}_{-0.0003}$  & 0.0175 $^{+0.0007}_{-0.0008}$  \\
      \midrule      
      Light BSM &  0.0037 $\pm$ 0.0004  & 0.0075 $\pm$ 0.0004  & 0.0211 $^{+0.0007}_{-0.0008}$ \\
      \midrule      
      Heavy BSM &  0.0048 $\pm$ 0.0004  & 0.0103 $\pm$ 0.0004  & 0.0242 $^{+0.0007}_{-0.0008}$ \\
      \bottomrule
    \end{tabular}
  }
  
  \vskip 0.5cm
  {\footnotesize
    \begin{tabular}{l c c c}
  \toprule
  & \multicolumn{3}{c}{$\pttt{}$ [\GeV{}]} \\
  $\AC{}$ &  $<25$ & $25$--$60$ & $>60$ \\
      \midrule      
      
            Data   & 0.044 $\pm$ 0.088 &  0.004 $\pm$ 0.066 &  0.002 $\pm$ 0.062 \\
      \midrule      
      SM     & 0.0141 $\pm$ 0.0007 & $-0.0051$ $\pm$ 0.0003 & $-0.0026$ $\pm$ 0.0002 \\
      \bottomrule
    \end{tabular}
}
\end{table}

\begin{table}[ht]
  \centering
    \caption{Correlation coefficients $\rho_{i,j}$ for the statistical and systematic uncertainties 
    between the $i$-th and $j$-th bin of the differential \AC{} 
  measurement as a function of the \ttbar{} invariant mass, \mtt{} (top), the \ttbar{} velocity along the z-axis, \betatt{} 
  (bottom left), and the transverse momentum, \pttt{} (bottom right). }
  
  \label{table:corr_diff}

  \begin{tabular}{l c c c c c c}
  \toprule
  $\rho_{ij}$& \multicolumn{6}{c}{$\mtt{}$ [\GeV{}]} \\
  $\mtt{}$ [\GeV{}] &  $<420$  &  $420$--$500$  &  $500$--$600$  &  $600$--$750$  &  $750$--$900$  &  $>900$ \\
      \midrule      
      $<420$     & 1. &  $-0.263$  &  0.076  & $-0.034$  & $-0.017$  &  $-0.001$ \\
      $420$--$500$  &    &   1.     & $-0.578$  &  0.195  & $-0.035$  &  $-0.002$ \\
      $500$--$600$  &    &          &  1.     & $-0.591$  &  0.160  &  $-0.028$ \\
      $600$--$750$  &    &          &         &  1.     & $-0.573$  &   0.132 \\
      $750$--$900$  &    &          &         &         &  1.     &  $-0.487$ \\
      $>900$     &    &          &         &         &         &   1.    \\
      
      \bottomrule
    \end{tabular}
      \vskip 0.5cm
\begin{minipage}{.45\textwidth}
      \begin{tabular}{l c c c}
  \toprule
  $\rho_{ij}$ & \multicolumn{3}{c}{$\betatt{}$} \\
   $\betatt{}$&  $<0.3$  &  $0.3$--$0.6$  &  $0.6$--$1.0$  \\
      \midrule      
      $<0.3$     & 1. &  $-0.262$  &  0.095 \\
      $0.3$--$0.6$  &    &   1.     & $-0.073$ \\
      $0.6$--$1.0$  &    &          &  1.    \\ 
      \bottomrule
    \end{tabular}
\end{minipage}\hfill
\begin{minipage}{.45\textwidth}
   \begin{tabular}{l c c c}
  \toprule
   $\rho_{ij}$ &\multicolumn{3}{c}{$\pttt{}$ [\GeV{}]}   \\
  $\pttt{}$ [\GeV{}] &  $<25$ & $25$--$60$ & $>60$ \\
      \midrule      
      
      $<25$   & 1. &  $-0.812$  &  0.431 \\
      $25$--$60$ &    &   1.     & $-0.722$ \\
      $>60$   &    &          &  1.    \\       
      \bottomrule
    \end{tabular}
\end{minipage}
\end{table}

\clearpage

\subsection{Interpretation}
\label{Interpretation}
\begin{figure}[!htb]
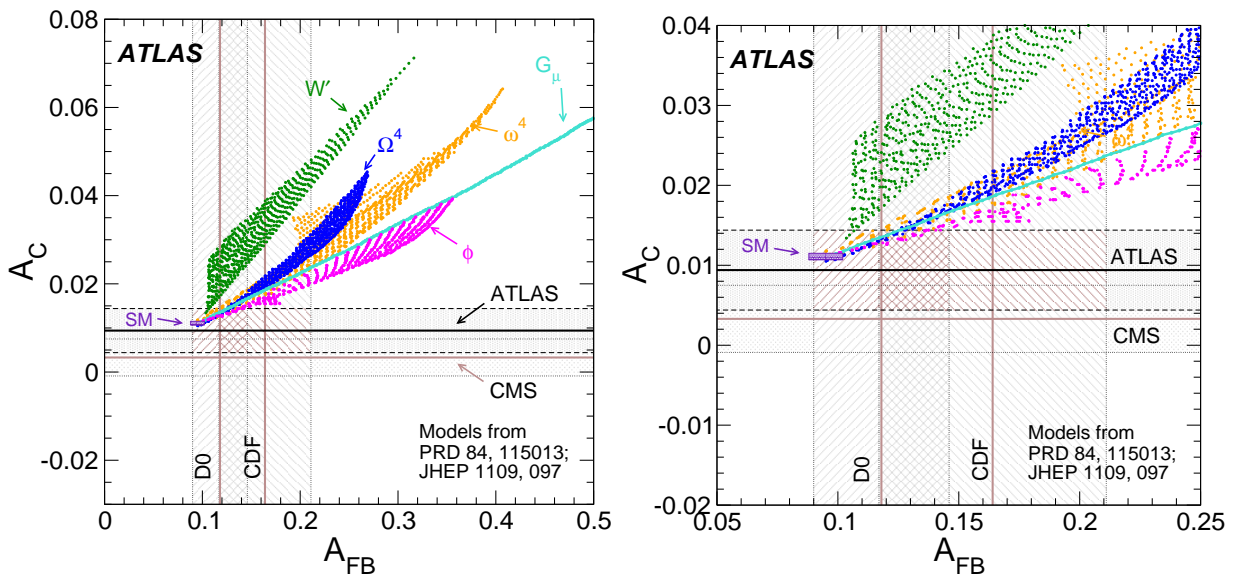
\centering
\includegraphics[trim={0 0 0 0.07cm},clip,width=0.495\textwidth]{acvsafb.eps} 
\includegraphics[trim={0 0 0 0.07cm},clip,width=0.495\textwidth]{acvsafb_zoom.eps} 
\caption{Measured inclusive charge asymmetries \AC{} at the LHC versus
forward--backward asymmetries $A_{\mathrm{FB}}$
at Tevatron, compared with the SM predictions~\cite{Czakon:2014xsa,Bern:2012} as well as predictions incorporating various potential 
BSM contributions~\cite{AguilarSaavedra:2011hz,AguilarSaavedra:2011ug}: a $W'$ boson, a heavy axigluon 
($\mathcal{G}_{\mu}$), a scalar isodoublet ($\phi$), a colour-triplet scalar ($\omega^4$), 
and a colour-sextet scalar ($\Omega^4$). The horizontal bands and lines correspond to the ATLAS and CMS measurements, while the vertical ones 
correspond to the CDF and D0 measurements. The uncertainty bands correspond to a 68\% confidence level interval.
The figure on the right is a zoomed-in version of the figure on the left.}
\label{fig:interpretation}
\end{figure}

Figure~\ref{fig:interpretation} shows the inclusive \AC{} measurement presented in Sect.~\ref{sec:result}. 
The measurement is compared to the \ttbar{} forward--backward asymmetry\footnote{
The \ttbar\ asymmetry at the Tevatron is measured as a forward--backward asymmetry and defined as
 $A_{\mathrm{FB}}= \frac{N(\Delta y>0)-N(\Delta y<0)}{N(\Delta y>0)+N(\Delta y<0)}$.} $A_{\mathrm{FB}}$
measured at the Tevatron by CDF and D0 experiments.
Predictions given by several BSM models, the details of which can be found 
in Refs.~\cite{AguilarSaavedra:2011hz,AguilarSaavedra:2011ug}, are also displayed.  These BSM models 
include a $W'$ boson, a heavy axigluon ($\mathcal{G}_{\mu}$), a scalar isodoublet ($\phi$), 
a colour-triplet scalar ($\omega^4$), and a colour-sextet scalar ($\Omega^4$).   
For each model, the predictions for $A_{\mathrm{FB}}$ and \AC{} are derived using the PROTOS
generator~\cite{AguilarSaavedra:2008gt} with the constraints described in Ref.~\cite{ATLAS_ljets}.
The ranges of predicted values for $A_{\mathrm{FB}}$  and \AC{}
for a given set of BSM model are also shown. The
BSM physics contributions are computed using the tree-level SM amplitude plus the one(s)
from the new particle(s),  to account for the interference between the two contributions.
The phase-space of the parameters describing the various BSM models (such as the BSM particle masses and couplings) 
is limited by the measurement presented in this paper.

%% file: summary.tex
\section{Conclusion}
\label{sec:conclusion}

The top-quark pair production charge asymmetry was measured with $pp$ collisions
at the LHC using an integrated luminosity of 20.3~\ifb{}
recorded by the ATLAS experiment at a centre-of-mass energy of
$\sqrt{s}=8$~\TeV{} in \ttbar{}
events with a single lepton (electron or muon), at
least four jets and large missing transverse momentum.
The reconstruction of \ttbar{} events was performed using a kinematic fit. 
The reconstructed inclusive distribution of \dy{} and the distributions as a function of \mtt{}, \pttt{} and \betatt{}
were unfolded to obtain results that can be directly
compared to theoretical computations.
The measured inclusive \ttbar{} production charge asymmetry is 
$ \AC{} = 0.009 \pm 0.005$ (stat.$+$syst.), to be compared to the
SM prediction $\AC{}=0.0111\pm0.0004$~\cite{Bern:2012}.
All measurements presented in this paper are statistically limited and are found
to be compatible with the SM prediction within the uncertainties. 
The precision of the measurements also allows for the exclusion of a large phase-space of the parameters describing various BSM models.

%% file: Acknowledgements.tex

We thank CERN for the very successful operation of the LHC, as well as the
support staff from our institutions without whom ATLAS could not be
operated efficiently.

We acknowledge the support of ANPCyT, Argentina; YerPhI, Armenia; ARC, Australia; BMWFW and FWF, Austria; ANAS, Azerbaijan; SSTC, Belarus; CNPq and FAPESP, Brazil; NSERC, NRC and CFI, Canada; CERN; CONICYT, Chile; CAS, MOST and NSFC, China; COLCIENCIAS, Colombia; MSMT CR, MPO CR and VSC CR, Czech Republic; DNRF, DNSRC and Lundbeck Foundation, Denmark; IN2P3-CNRS, CEA-DSM/IRFU, France; GNSF, Georgia; BMBF, HGF, and MPG, Germany; GSRT, Greece; RGC, Hong Kong SAR, China; ISF, I-CORE and Benoziyo Center, Israel; INFN, Italy; MEXT and JSPS, Japan; CNRST, Morocco; FOM and NWO, Netherlands; RCN, Norway; MNiSW and NCN, Poland; FCT, Portugal; MNE/IFA, Romania; MES of Russia and NRC KI, Russian Federation; JINR; MESTD, Serbia; MSSR, Slovakia; ARRS and MIZ\v{S}, Slovenia; DST/NRF, South Africa; MINECO, Spain; SRC and Wallenberg Foundation, Sweden; SERI, SNSF and Cantons of Bern and Geneva, Switzerland; MOST, Taiwan; TAEK, Turkey; STFC, United Kingdom; DOE and NSF, United States of America. In addition, individual groups and members have received support from BCKDF, the Canada Council, CANARIE, CRC, Compute Canada, FQRNT, and the Ontario Innovation Trust, Canada; EPLANET, ERC, FP7, Horizon 2020 and Marie Skłodowska-Curie Actions, European Union; Investissements d'Avenir Labex and Idex, ANR, Region Auvergne and Fondation Partager le Savoir, France; DFG and AvH Foundation, Germany; Herakleitos, Thales and Aristeia programmes co-financed by EU-ESF and the Greek NSRF; BSF, GIF and Minerva, Israel; BRF, Norway; the Royal Society and Leverhulme Trust, United Kingdom.

The crucial computing support from all WLCG partners is acknowledged
gratefully, in particular from CERN and the ATLAS Tier-1 facilities at
TRIUMF (Canada), NDGF (Denmark, Norway, Sweden), CC-IN2P3 (France),
KIT/GridKA (Germany), INFN-CNAF (Italy), NL-T1 (Netherlands), PIC (Spain),
ASGC (Taiwan), RAL (UK) and BNL (USA) and in the Tier-2 facilities
worldwide.

%% file: atlas_authlist.tex
\begin{flushleft}
{\Large The ATLAS Collaboration}

\bigskip

G.~Aad$^{\textrm 85}$,
B.~Abbott$^{\textrm 113}$,
J.~Abdallah$^{\textrm 151}$,
O.~Abdinov$^{\textrm 11}$,
R.~Aben$^{\textrm 107}$,
M.~Abolins$^{\textrm 90}$,
O.S.~AbouZeid$^{\textrm 158}$,
H.~Abramowicz$^{\textrm 153}$,
H.~Abreu$^{\textrm 152}$,
R.~Abreu$^{\textrm 116}$
% ...
\bigskip
\\
% ...
$^{11}$ Institute of Physics, Azerbaijan Academy of Sciences, Baku, Azerbaijan\\
% ...
$^{85}$ CPPM, Aix-Marseille Universit{\'e} and CNRS/IN2P3, Marseille, France\\
% ...
$^{90}$ Department of Physics and Astronomy, Michigan State University, East Lansing MI, United States of America\\
% ...
$^{107}$ Nikhef National Institute for Subatomic Physics and University of Amsterdam, Amsterdam, Netherlands\\
% ...
$^{113}$ Homer L. Dodge Department of Physics and Astronomy, University of Oklahoma, Norman OK, United States of America\\
% ...
$^{116}$ Center for High Energy Physics, University of Oregon, Eugene OR, United States of America\\
% ...
$^{151}$ Institute of Physics, Academia Sinica, Taipei, Taiwan\\
$^{152}$ Department of Physics, Technion: Israel Institute of Technology, Haifa, Israel\\
$^{153}$ Raymond and Beverly Sackler School of Physics and Astronomy, Tel Aviv University, Tel Aviv, Israel\\
% ...
$^{158}$ Department of Physics, University of Toronto, Toronto ON, Canada\\
% ...
\end{flushleft}
